\documentclass[twocolumn,prb]{revtex4}

\usepackage{graphicx,amssymb,soul,color,amsmath,bm}
\usepackage{epstopdf,hyperref}
\usepackage{braket,dsfont}
\usepackage[mathcal]{euscript}

\hypersetup{colorlinks=true,citecolor=blue,linkcolor=magenta}

\sethlcolor{yellow}
\allowdisplaybreaks

\newcommand{\refeq}[1]{(\ref{#1})}

\begin{document}

\title{Time and momentum-resolved tunneling spectroscopy of pump-driven non-thermal excitations in Mott insulators}
\author{Krissia Zawadzki}
\affiliation{Department of Physics, Northeastern University, Boston, Massachusetts 02115, USA}
\author{Adrian E. Feiguin}
\affiliation{Department of Physics, Northeastern University, Boston, Massachusetts 02115, USA}

\date{\today}
\begin{abstract}
We present a computational technique to calculate time and momentum resolved non-equilibrium spectral density of correlated systems using a tunneling approach akin scanning tunneling spectroscopy. The important difference is that our probe is extended, basically a copy of the sample, allowing one to extract the momentum information of the excitations. We illustrate the method by measuring the spectrum of a Mott-insulating extended Hubbard chain after a sudden quench with the aid of time-dependent density matrix renormalization group (tDMRG) calculations. We demonstrate that the system realizes a non-thermal state that is an admixture of spin and charge density wave states, with corresponding signatures that are recognizable as in-gap sub-bands. In particular, we identify a band of excitons and one of stable anti-bound states at high energies that gains enhanced visibility after the pump. We do not appreciate noticeable relaxation within the time-scales considered, which is attributed to the lack of decay channels due to spin-charge separation. These ideas can be readily applied to study transient dynamics and spectral signatures of correlation-driven non-equilibrium processes.
\end{abstract}

\maketitle

\section{Introduction}
With the advent of new powerful light sources, experimentalists can shake the excitations of a system and probe states present in the spectrum that are not accessible via finite-temperature measurements. By means of ultrafast light pulses, electrons can be excited above any intrinsic energy scale, and the competition between different degrees of freedom can be manipulated. \cite{Cavalieri2007,Orenstein2012c,Chollet2005,Corkum2007a,Onda2008}. 
The resulting non-thermal states after photoexcitation often contain coexisting orders that are not usually present in the ground or thermal states\cite{Orenstein2012c,Kampfrath2013}.
This new knob can be used to stabilize ``hidden'' phases that reside at higher energies, such as  superconductivity\cite{Fausti2011}, and to induce or disrupt charge, magnetic, or orbital order \cite{Polli2007,Ehrke2011,Zhang2016,Casals2016,Chollet2005,Onda2008, Okamoto2007b}. 

Time-resolved femtosecond photoemission spectroscopy has been one of the most a used techniques to monitor in real time and with atomic resolution the ultrafast quasiparticle dynamics in correlated-electron materials
\cite{Plummer1997,ramakrishna2001,domcke1991,sebastian2006}. The experimental protocol  \cite{Cavalieri2007,Smallwood2016} starts with an intense pulse of radiation that `pumps' the
system into a highly excited non-equilibrium state. After a
variable time delay, the system is subject to a weak
probe pulse of higher energy photons, ejecting photoelectrons
which are detected with energy (and angle)
resolution.
By means of this powerful tool, one can peek into the different decay mechanisms taking place, and experimentally unveil the complex and rich interplay between charge, spin, orbital and vibrational degrees of freedom.

Notwithstanding, theoretically reproducing time- and angle-resolved photoemission spectra is computationally challenging and expensive. It can be numerically carried out only in small systems, as it requires the full knowledge of the eigenstates and the calculation of a two-time correlator \cite{Freericks2009,Shao2016}. In the equilibrium steady state, approximations can be made by using the single-particle Green's function, but all information about transient and the actual decay mechanisms during the relaxation process is lost.
In a non-thermal state far from equilibrium, the imaginary part of the equilibrium retarded Green's function $G(\omega)$ is not guaranteed to be positive and does not yield meaningful information about the orbital occupation (it is not a density of states).

\begin{figure}
\centering
\includegraphics[width=0.45 \textwidth]{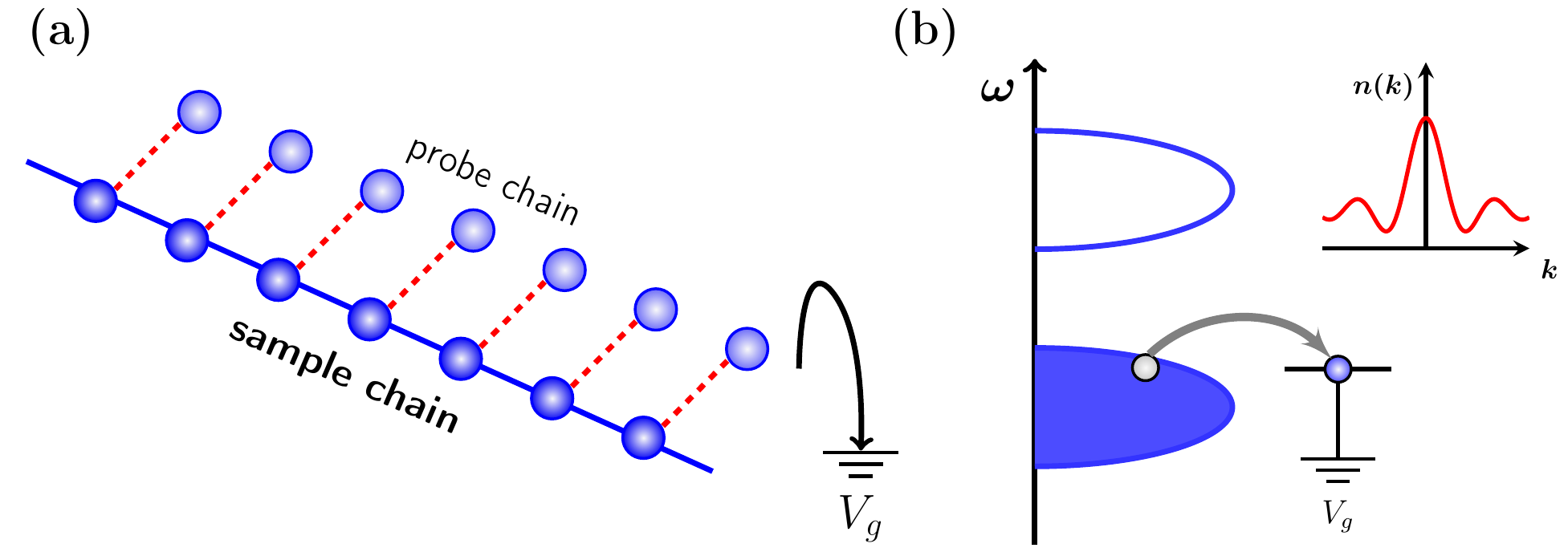}
\caption{(a) Proposed tunneling setup: a sample chain is connected to a probe chain via a tunneling barrier. The probe chain is set at a target gate voltage $V_g$. (b) Particles tunnel for a short period of time to the probe chain, where their momentum distribution $n(k)$ is measured. 
}
\label{fig:tunnel}
\end{figure}

We hereby propose a different approach to investigate these quantities using a tunneling technique.
We focus on a geometry that was first suggested in Ref.\onlinecite{Carpentier2002}, and later realized experimentally in Refs.\onlinecite{Auslaender2002,Auslaender2005} for conducting momentum-resolved tunneling spectroscopies on one-dimensional (1D) systems. Unlike scanning tunneling spectroscopy, where the probe yields only local information\cite{Cohen2014}, an extended one-dimensional wire can provide momentum resolution. Electrons can tunnel from the sample into the one-dimensional non-interacting lead that is placed parallel to it. Since this occurs in the transverse direction, momentum conservation along the probe direction is ensured. A gate voltage $V_g$ is applied to the probe wire and energy conservation implies that only electrons with energy $\omega=V_g$ can tunnel. Momentum resolution is achieved through the application of a magnetic field perpendicular to the plane of the sample and probe wires. A similar scheme was recently proposed for performing momentum-resolved spectroscopies on cold atomic systems: instead of a voltage, an RF field or the shaking of the lattice can yield transitions at a target frequency\cite{Kantian2015, Bohrdt2017}. In this variation, as particles tunnel to the second channel, momentum is mapped via time of flight.

In section \ref{method} we describe in detail the implementation and illustrate with simple examples. In section \ref{results} we present results for an interacting system --the extended Hubbard model-- after a quench, and we close with a summary and discussion.  

\section{Method} \label{method}
We propose to computationally carry out a hybrid method combining ideas from the aforementioned setups: after the system has been photoexcited, we allow for electrons to tunnel into an empty parallel wire which has been set at a given gate voltage, as shown in Fig.\ref{fig:tunnel}(a). Only electrons at a particular energy $V_g$ can tunnel, and we can then access the occupation of each state with momentum resolution by simply calculating the momentum distribution function of the probe wire.

\subsection{Non-interacting fermions}
We illustrate this idea with the simple example of non-interacting fermions, whose Hamiltonian reads 
\begin{equation}
H_0=-J \sum_{i=1,\sigma}^{L-1} \left(c^\dagger_{i} c_{i+1}+\mathrm{h.c.}\right) = \sum_k \omega_k c^\dagger_kc_k,
\label{h0}
\end{equation} 
where $c^\dagger_{i}$ and $c_i$ are the usual creation and annihilation fermion operators (we ignore the spin index for now) and $w_k=-2J\cos{k}$. We take the inter-atomic distance as unity and we express all energies in units of the hopping parameter $J$ (the symbol ``$t$'' will be reserved to represent time, which will be expressed in units of $1/J$). 

\onecolumngrid

\begin{figure}[t]
\centering
\includegraphics[width=\textwidth]{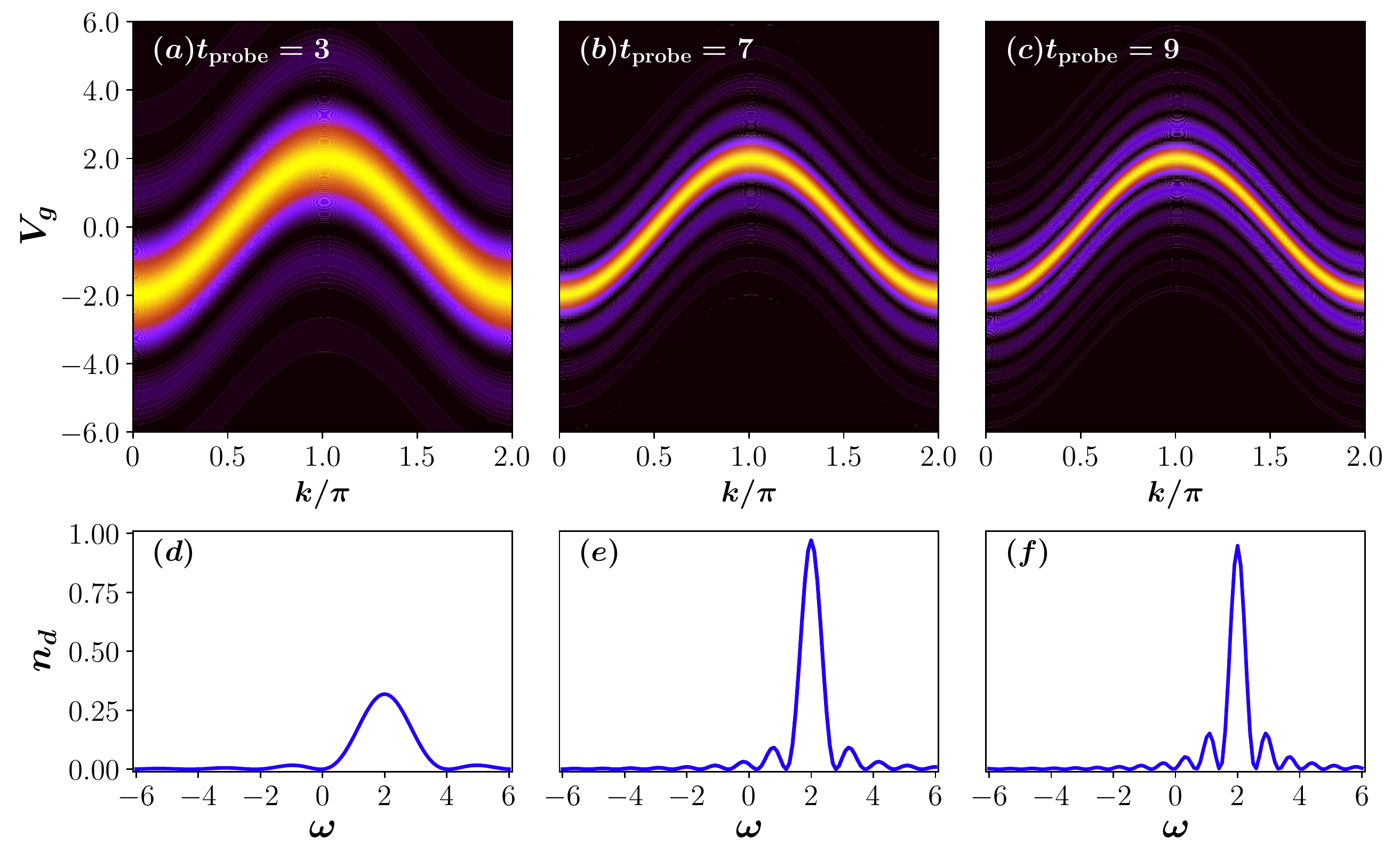}
\caption{Spectral function for a tight-binding chain obtained from the momentum distribution function $n_d(k)$ of a parallel probing chain after being in contact with the physical system for a time $t_{probe}$. Each momentum $k$ is obtained exactly by solving the two-level problem described above. Panels (a),(b),(c) correspond to $t_{probe}=3,7,9$ in units of $1/J$. Curves (d),(e),(f) show a cut in frequency along the $k=\pi$ line for the same times as in (a),(b) and (c).
}
\label{fig:tb}
\end{figure}
\twocolumngrid

A second ``probe'' chain is included as
\begin{equation}
H_{\mathrm{probe}}= V_g \sum_{i=1}^L d^\dagger_i d_i = V_g \sum_k d^\dagger_k d_k,
\label{hprobe}
\end{equation}
where we distinguish the operators $d^\dagger$ and $d$ acting on the probe.
Notice that there is no hopping nor interactions along the probe chain: it consists of isolated empty orbitals with a gate voltage (or chemical potential) $V_g$. At time $t=0$ the system is in the ground state of the physical chain at a fixed given density, while the probe chain is empty (this is ensured by initially setting $V_g$ to a very large positive value). Then, both chains are connected by means of a tunneling term:
\begin{equation}
H_\mathrm{tunnel}= J' \sum_{i=1}^L \left( d^\dagger_i c_i + \mathrm{h.c.}\right) = J' \sum_k \left( d^\dagger_k c_k + \mathrm{h.c.}\right).
\label{htunnel}
\end{equation}
Putting together Eqs.\refeq{h0},\refeq{hprobe} and \refeq{htunnel}, the full problem becomes the sum of $L$ independent tunneling terms, which can be readily solved.
For simplicity, we look at an eigenstate at temperature $T=0$, in which a single particle orbital with momentum $k$ is either empty or occupied. 
The single particle states are $|n(k),n_d(k)\rangle = {|1,0\rangle; |0,1\rangle}$ where $1$ and $0$ represent the occupancy of the physical orbital or the probe orbital with momentum $k$.
The Hamiltonian for the two level system is :
$$ H=\begin{bmatrix} \omega_k & -J' \\ -J' & V_g \end{bmatrix}
$$
with ground state energies
$$ E_\pm(k) = \frac{V_g+\omega_k}{2}\pm\sqrt{\left(\frac{V_g-\omega_k}{2}\right)^2+J^2},
$$
Starting from a initial state $|1,0\rangle$, the probability that a fermion is transferred to the corresponding empty probe state $k$ at time $t$ is simply $n_d(t)=|\langle t|0,1\rangle|^2 = 4A^2\sin^2{\left(\frac{E_+-E_-}{2}t\right)}$, with $A=J'W/(J'+W)^2, W=\omega_k-E_+$.
This function oscillates in time with a period $\tau=\pi/J'$ for $V_g=\omega_k$. In order to maximize the ``visibility'' one needs to measure the density of the probe state $k$ at time $t_{\mathrm{max}}=\pi/2J'$. 
As a function of $V_g$, the probability is peaked at $V_g=\omega_k$ (with smaller satellite peaks), and its width gets narrower as $J'\rightarrow 0$, or $t_\mathrm{max}\rightarrow \infty$, as shown in Fig.\ref{fig:tb}. This is nothing else but Fermi's golden rule and a manifestation of the uncertainty principle: to obtain sharper resolution in energy, one needs to choose a small coupling $J'$ and measure at very long times.

\subsection{General formulation away from equilibrium}

We now consider the full many-body Hamiltonian $H$ and a generic initial state  $\rho_0 = \rho \otimes |0\rangle\langle0|$. The density matrix $\rho$ represents the state of the many-body system: $\rho=\sum_{n,n'} a_{n'}^*a_n |n'\rangle\langle n|$, where the states $|n\rangle$ are the eigenstates of $H$. We point out that this is a general scenario, in which the system may have been driven away from equilibrium by an external perturbation or a quench and $H$ is the final Hamiltonian; the equilibrium case is simply recovered by taking a diagonal density matrix. We assume that the measurement process starts suddenly some time after the perturbation, which for simplicity of notation we label as $t=0$, and the system evolves thereafter under the action of a time-independent Hamiltonian ({\it i.e.},any time-dependence in the Hamiltonian is ``frozen'' at time $t=0$). Following closely the discussion in Ref.\onlinecite{Kantian2015} we find that in second order of perturbation, the occupation of the state $k$ of the probe system is given as (we ignore the spin index for now):
\begin{equation}
    \langle n_d(k,t)\rangle = \int_0^t dt_1 \int_0^t dt_2 \langle V_k(t_1) n_d(k) V_k(t_2) \rangle,
\end{equation}
where $V_k = -J'\left(d^\dagger_k c_k +  \mathrm{h.c.}\right)$. 
The averages are with respect to the initial state $\langle \cdots \rangle = \mathrm{Tr}(\cdots \rho_0)$ and we work in the interaction picture $O(t)=e^{i(H+V_gn_d)t} Oe^{-i(H+V_gn_dt)t}$ (from now on we ignore the subindex $k$ for convenience).
Since the probe orbital is initially empty, the only term surviving in this expression is:
\begin{eqnarray}
 \langle n_d\rangle & = &     
J'^2 \int_0^t dt_1 \int_0^t dt_2 
\langle c^\dagger(t_1) d(t_1) n_d d^\dagger(t_2) c(t_2) \rangle. 
\label{integral}
\end{eqnarray}
Moreover, noticing that the initial state is a product state, we can readily evaluate the contribution of the probe orbital to this expression: $\mathrm{Tr}_{probe}(d(t_1)n_d d^\dagger(t_2)|0\rangle\langle 0|) = e^{iV_g(t_2-t_1)}$. Explicitly, Eq.(\ref{integral}) becomes: 
\begin{widetext}
\begin{eqnarray}
\langle n_d\rangle &=& 
J'^2  \int_0^t dt_1 \int_0^t dt_2 
e^{iV_g(t_2-t_1)} \langle c^\dagger(t_1)c(t_2) \rangle \nonumber \\
& = & J'^2  \int_0^t dt_1 \int_0^t dt_2 
e^{iV_g(t_2-t_1)} \sum_{n,n'} a_{n'}^*a_n e^{i(E_{n'}t_1-E_nt_2)}\langle n'|c^\dagger e^{iH(t_2-t_1)}c|n\rangle \nonumber \\
& = & J'^2  \int_0^t dt_1 \int_0^t dt_2 
e^{iV_g(t_2-t_1)} \sum_{n,n'}\sum_{m} a_{n'}^*a_n e^{i(E_{n'}-E_m)t_1}e^{i(E_m-E_n)t_2}\langle n'|c^\dagger |m\rangle \langle m|c|n\rangle \nonumber \\
& = & J'^2    
 \sum_{m} \left(\int_0^t dt_1 \sum_{n'} a_{n'}^* e^{-i(E_m-E_{n'}+V_g)t_1}\langle n'|c^\dagger|m\rangle \right) \times \left(\int_0^t dt_2 \sum_n a_n e^{i(E_m-E_n+V_g)t_2}\langle m|c|n\rangle \right) \nonumber  \\
 & = & J'^2 \sum_m \left| \sum_n a_n \frac{\sin{[(V_g-\omega_{nm})t/2]}}{(V_g-\omega_{nm})/2} \langle m|c|n\rangle \right|^2, 
\end{eqnarray}
\end{widetext}
with $\omega_{nm}=E_n(N)-E_m(N-1)$, $N$ being the number of particles. 
Hence, the resulting occupation will be peaked at the gate voltages corresponding to the allowable transitions, $\langle m(N-1)|c_k|n(N)\rangle$ , weighed by the initial occupation of the eigenstates. We notice that the argument in the integral is just the lesser Green's function $G^<(t_1,t_2)=\langle c^\dagger(t_1)c(t_2)\rangle$ and this equation is identical to the one derived in Ref.\onlinecite{Freericks2009} to describe a time-resolved photoemission experiment. 

For large $t$, the quantity $n_d(k,t)$ converges to a sum of Dirac deltas and yields an expression proportional to the system's spectral function. Clearly, at long times the electron will be reflected and tunnel back to the system so, in reality, to improve the energy resolution one needs to pick $J'$ small. In general, we take as a rule of thumb $t_{\mathrm{max}}=\pi/2J'$ in all cases. We point out that a time-dependent tunneling term could also be considered, which translates into the introduction of an envelope function, as done in Ref.\onlinecite{Freericks2009}.

\section{Results}\label{results}
We now demonstrate an application of this scheme to explore competing orders and excitations in one-dimensional correlated materials. It is known that in 1D systems, the band edge singularity could give rise to a high-differential optical gain, with potential applications such as light-emitting diodes, lasers, sensors, and molecular switches \cite{brown1995,burroughes1990,dodabalapur1995,dodabalapur1995b,hide1996,yang1998,schmitz2001,nitzan2003}. There is great deal of interest in the optical properties of 1D materials in the presence of correlations, when a gap arises as a result of electronic interactions. Moreover, the emergence of excitonic excitations, has been subject of attention of a number of theoretical\cite{jeckelmann2000,tsutsui2000,essler2001,jeckelmann2003,gallagher1997,barford2002,Gebhard1997,Kancharla2001,Mizuno2000,Glocke2007,Matsueda2008,Lu2015,exciton1,exciton2,Rincon2014}and experimental \cite{ono2005,Schlappa2012} works.

The minimal model to study correlated polymers is the so-called ``$U-V$'' extended Hubbard model:
\begin{eqnarray}
H&=&-J \sum_{i=1,\sigma}^{L-1} \left(c^\dagger_{i\sigma} c_{i+1\sigma}+\mathrm{h.c.}\right) + \nonumber \\
&+&U \sum_{i=1}^L 
\left(n_{i\uparrow}-\frac{1}{2}\right)\left(n_{i\downarrow}-\frac{1}{2}\right)+  \\ 
&+& V\sum_{i=1}^{L-1} \left(n_{i}-1\right)\left(n_{i+1}-1\right). \nonumber
\label{Hubbard}
\end{eqnarray}
 Here, $c^\dagger_{i\sigma}$ creates an electron of spin $\sigma$ on the
$i^{\rm th}$ site along a chain of length $L$. The on-site and nearest-neighbor Coulomb repulsion are parametrized by $U$ and $V$, respectively. 

The physics of one-dimensional strongly correlated fermionic systems can generally be described in terms of Luttinger liquid theory.
In a Luttinger liquid (LL) \cite{Haldane1981,GogolinBook,GiamarchiBook}, the natural
excitations are collective density fluctuations, that carry either spin
(``spinons''), or charge (``holons'').
 This leads to the spin-charge separation picture, in
which a fermion injected into the system breaks down into excitations, each with a characteristic energy scale and velocity (one for the charge, one for the spin).
Spin-charge separation acts as a constraint for the dynamics of the system, that cannot relax to a thermal state after a quench or non-equilibrium situation. The lack of thermalization implies that it might be possible to `trap' the system in an excited state for very long times.

\begin{figure}
\centering
\includegraphics[width=0.48 \textwidth]{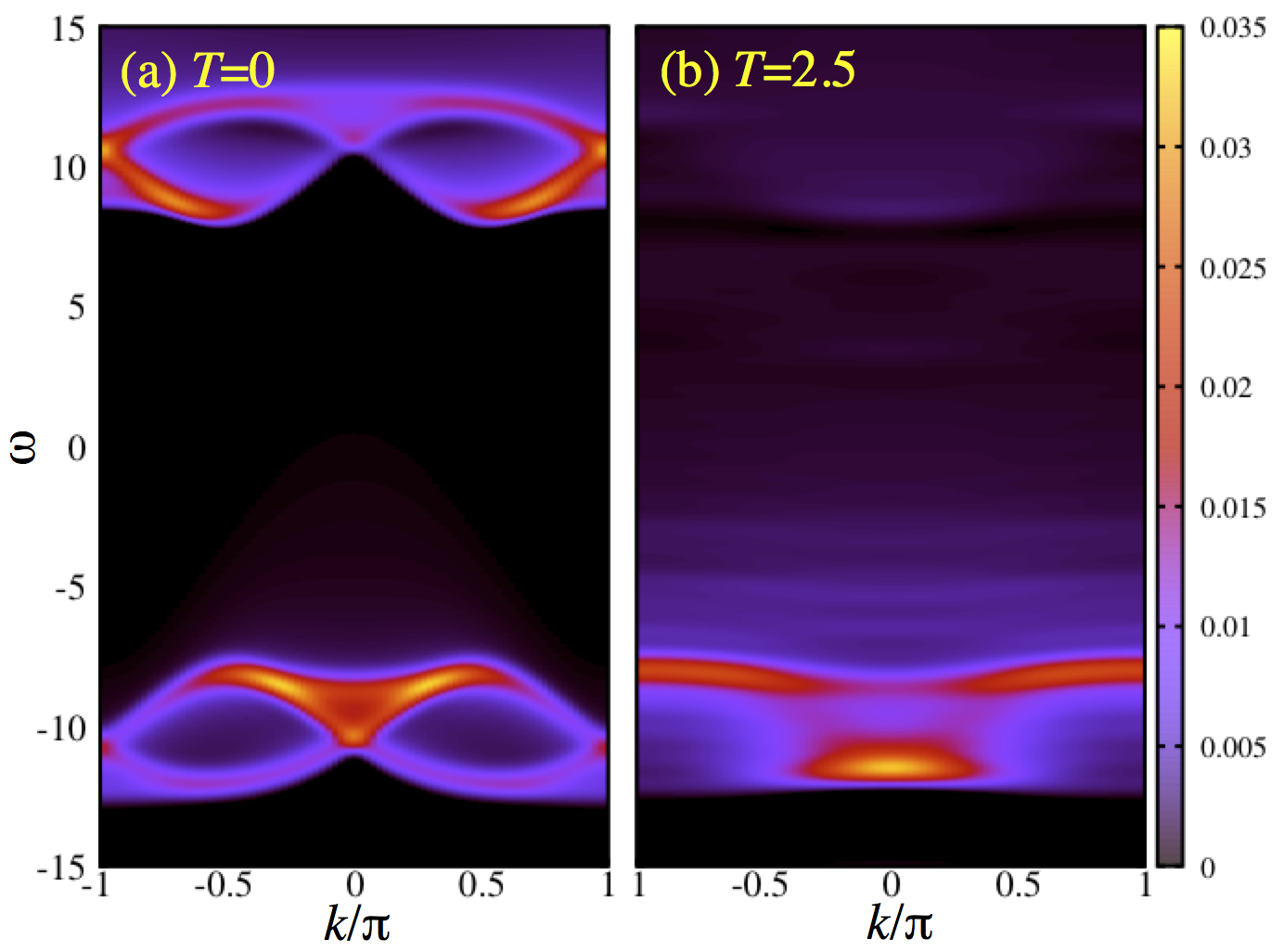}
\caption{Momentum resolved spectrum of the 1D extended Hubbard model at half-filling at (a) zero temperature, where negative(positive) frequencies correspond to occupied(empty) states, and (b) $T=2.5J$, obtained with the tDMRG method for a chain of length $L=40$, and interaction $U=20, V=5$.
}
\label{fig:arpes}
\end{figure}

As a proof of concept we conduct a numerical experiment using the time-dependent density matrix renormalization group method (tDMRG) \cite{White2004a,Daley2004,vietri,Paeckel2019} on chains of length $L=32$ and with parameters $U=20$ and $V=5$. This choice may seem exaggerated, but is justified: it will provide us with a large Mott gap $\Delta \sim U-V$, and allow us to resolve any features that may appear inside the gap with more detail and well separated from the bands. The ground state of the system at half-filling is a Mott insulator with dominant power-law decaying quasi-long-range antiferromagnetic order, or SDW phase \cite{Nakamura2000,Jeckelmann2002,Sandvik2003,Tsuchiizu2004}. The optical conductivity and Raman spectrum reveal the existence of sharp excitonic peaks with a weak continuous band of free excitations of width $\sim 8J$ \cite{jeckelmann2000,tsutsui2000,essler2001,jeckelmann2003,gallagher1997,Gebhard1997,Kancharla2001}. 
However, these optical excitations are not present in the spectrum, shown in Fig.\ref{fig:arpes}(a) as a reference, also obtained using tDMRG with $m=600$ states. The lower and upper Hubbard bands are well separated from each other by a wide Mott gap, and no remarkable features are observed, besides the characteristic holon and spinon dispersions.

\begin{figure}
\centering
\includegraphics[width=0.48 \textwidth]{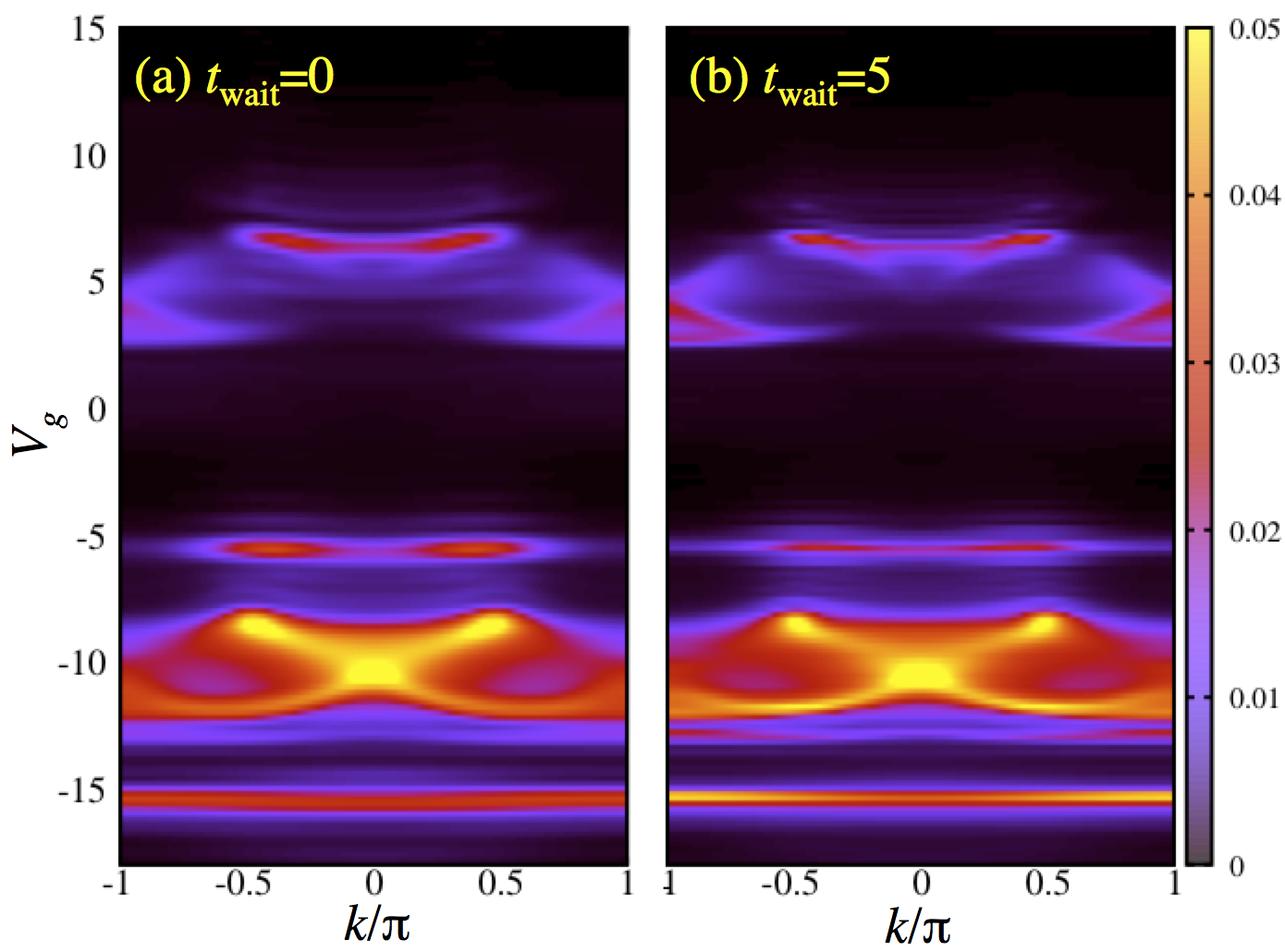}
\caption{Momentum resolved tunneling spectrum of the 1D extended Hubbard model at half-filling after a sudden quench in the interactions from $U_0=2,V_0=0$ to $U=20,V=5$, obtained with the tDMRG method for a chain of length $L=32$ a time (a) $t_\mathrm{wait}=0$ and (b) $t_\mathrm{wait}=5$ after the quench, and a probe time $t_\mathrm{probe}=7.9$}
\label{fig:tunnel_DMRG}
\end{figure}

To avoid considerations concerning pulse shape, frequency, and length, we simplify the discussion to the case of a quench, in which the system is prepared in the Mott insulating ground state of a system of $N=L$ electrons with $U_0=2, V_0=0$, and the interactions are suddenly changed to $U=20, V=5$. As a consequence, the final state will be a superposition of eigenstates that will exhibit a large number free holes and doublons, as well as excitons, occupying broad range of energies. 
In Fig.\ref{fig:tunnel_DMRG} we show results obtained using tunneling spectroscopy
 right after the quench. The probe is connected to the chain at time $t_\mathrm{wait}$ after the quench, and we plot the momentum distribution function of the probe chain at time $t_\mathrm{wait}+t_\mathrm{probe}$ as a function of momentum and gate voltage:
 \[
n_{d\sigma}(k)=\frac{1}{L}\sum_{j,l=1}^L e^{ik(j-l)}\langle d^\dagger_{j\sigma} d_{l\sigma}\rangle.
\]
 Notice that we use open boundary conditions throughout, which translates into some uncertainty in momentum. We scanned $V_g$ in steps of 0.2, implying 175 independent tDMRG simulations for each value of $t_\mathrm{wait}$. We took $J'=0.2$ and  used $m=200$ DMRG states, which yields a truncation error of the order of $10^{-4}$ in the worse cases.

\begin{figure}
\centering
\includegraphics[width=0.48 \textwidth]{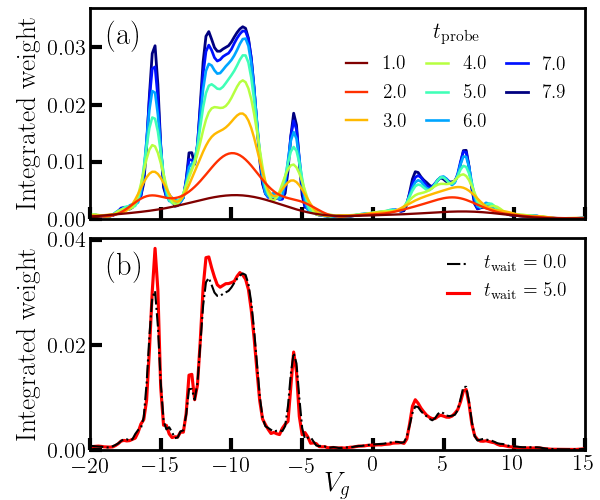}
\caption{(a) Integrated spectral weight as a function of $V_g$ for $t_\mathrm{wait}=0$ and different probe times, demonstrating the resolution improvement. (b) Same as (a) but for $t_\mathrm{wait}=0$ and 5, at the final $t_\mathrm{probe}=7.9$.}
\label{fig:integrated}
\end{figure}

As shown in Fig.\ref{fig:tunnel_DMRG}, besides some sharper and better defined features, we are not able to resolve a noticeable difference between the measurements right after the quench and at $t_\mathrm{wait}=5$. This is also reflected in the integrated weight over momenta, displayed in Fig.\ref{fig:integrated}: panel (a) illustrates how the visibility improves as a function of $t_\mathrm{probe}$  (see animations in the supplementary material), while in (b) we compare the two waiting times. In this case, we are able to resolve some minor differences that stem from the relaxation of excitations within the lower Hubbard band, indicating the lack of available channels for non-radiative decay or recombination. It is possible that these excitations cannot decay due to the energy mismatch between the bandwidth $W$ and the interactions, or thermalization occurs in timescales that far exceed the simulation time. This
bottleneck exists already in higher dimensions \cite{Sensarma2010, Eckstein2011, Lenarcic2013,Eckstein2016}. If the bandwidth is small, the number of available decay channels gets suppressed. However, in our case the in-gap states are not too far from each other, nor from the lower Hubbard band. In higher dimensions it was observed that the spin excitations are highly relevant for thermalization. It is possible that  spin-charge separation, which is more dramatic at large values of $U$, and the flat spinon dispersion for large values of $U$ do not allow for a wide range of energy and momenta for scattering.

The spectrum is very well resolved and displays many non-trivial features that are not present neither in the zero temperature spectrum nor the optical conductivity. In order to account for these results, we first assume the possibility that the system is in a thermal state. We have calculated the spectra for a wide range of temperature scales and have found that the final state after the quench does not correspond to a thermal distribution. For illustration, we display finite-temperature tDMRG\cite{Feiguin2005a} results at $T=2.5J$ in Fig.\ref{fig:arpes}(b). The first remarkable and most obvious feature of the spectrum is recognizable in the lower Hubbard band, which displays a dispersion rather resembling a tight-binding band of spinless fermions than the usual characteristics of fractionalized excitations seen in panel (a). This is actually expected, since in this regime the spin is completely incoherent (We refer the reader to Refs.\onlinecite{Cheianov2004,Cheianov2005,Abendschein2006,Fiete2007b,Halperin2007,Feiguin2009,Feiguin2011,Soltanieh-ha2014,Nocera2018} for a discussion of the finite-temperature spectra of 1D correlated systems). 
Moreover, we distinguish a distribution of spectral weight inside the gap due to the correlated nature of the problem \cite{Nocera2018}, a phenomenon that has been experimentally observed in the photoemission spectrum of the single chain Mott insulators Sr$_2$CuO$_2$\cite{kidd08} and Na$_{0.96}$V$_2$O$_5$ \cite{maekawa99}.  
On the other hand, the tunneling spectrum displays a quite large spectral weight inside the gap and in the upper Hubbard band, implying that if we had to assign a temperature to the system after the quench, it would have to be larger than the Mott gap. However, unlike the finite temperature case, the spinon and holon bands remain coherent. 

\begin{figure}
\centering
\includegraphics[width=0.48 \textwidth]{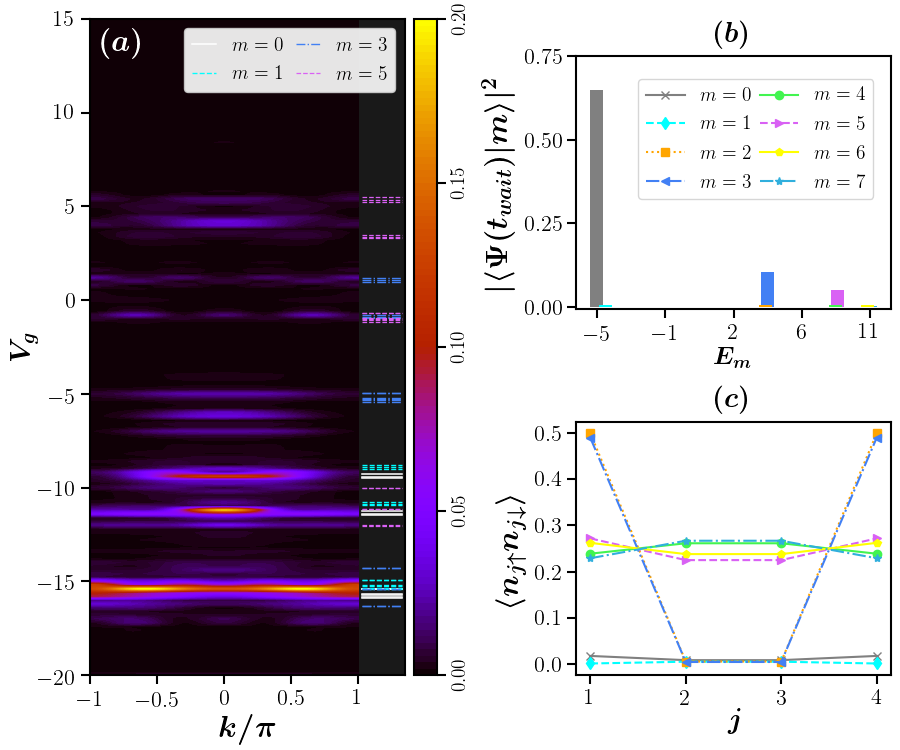}
\caption{(a) Same as Fig.\ref{fig:tunnel_DMRG} obtained with exact diagonalization for a chain with $L=4$ and 4 probe sites. Horizontal color bars represent different transition frequencies, as explained in the text. (b) Histogram showing the contribution of different eigenstates to the resulting distribution after the quench. (c) Local density of double occupied sites for each eigenstate.  }
\label{fig:tunnel_ED}
\end{figure}

In order to make sense of the unexpected features in the tunneling results, we carry out a similar simulation using exact diagonalization on a chain with $L=4$ sites with a parallel chain as a probe. The complexity of the problem is similar to that of a $2\times4$ Hubbard ladder with 4 electrons. Even though it is a small system and is likely very affected by boundary effects, it provides valuable intuition to interpret the tDMRG results. Following a similar protocol, we first resolve the tunneling spectrum, shown in Fig.\ref{fig:tunnel_ED}(a). Since we have access to all eigenstates and eigenvalues, we calculate all possible single particle excitation energies as $\omega_{nm} = E_n(N=L-1,S^z=1)-E_m(N=L,S^z=0)$, some of which are shown in the plot with different colors. The final state is predominantly a superposition of the ground state --which has dominant SDW correlations-- and two excited states, labelled $|m=3\rangle$ and $|m=5\rangle$ in Fig.\ref{fig:tunnel_ED}(b), that display CDW correlations, as shown in panel Fig.\ref{fig:tunnel_ED}(c). This enhancement of the charge order was previously observed in Ref.\onlinecite{Lu2012a} under the action of a driving field. We focus on the dominant features of the spectrum, namely, the flat bands at energy $\omega \sim -5$ and $\omega \sim -15$, and the in-gap spectral weight at energies between $\omega=5$ and $\omega=7$. The first one corresponds to breaking a holon-doublon pair on top of $|m=3\rangle$, while the in-gap weight corresponds to excitations on top of $|m=5\rangle$. The flat band  at high energies below the Fermi level is an excitation on top of the ground state that acquires an enhanced spectral weight after the pump. This high energy feature has been overlooked in prior studies of the model due to its very weak spectral signatures at zero-temperature, and indicates the presence of stable anti-bound states outside of the continuum.

 
\section{Conclusions}
To summarize, we have introduced a computational tunneling approach that allows one to access the time and momentum resolved spectrum of strongly correlated systems away from equilibrium, which previously could only be obtained from small systems with exact diagonalization. The formulation is general and does not depend on how the system is driven out of equilibrium. We have applied the method to study the dynamics of Mott insulating Hubbard chains after a quench and have been able to identify features in the spectrum corresponding to an admixture of SDW and CDW states, with a band of doublon-holon excitons and high-energy anti-bound states. This extremely powerful technique can be readily extended to arbitrary models under a variety of scenarios, giving access to transient dynamics and the ability to identify correlation-driven non-equilibrium processes behind pump-driven phase transitions and exciton decay and recombination. 


\acknowledgements
We thank A. Nocera for valuable comments. We acknowledge generous computational resources provided by Northeastern University's Discovery Cluster at the Massachusetts Green High Performance Computing Center (MGHPCC). KZ is supported by a Faculty of the Future fellowship of the Schlumberger Foundation. AEF acknowledges the U.S. Department of Energy, Office of Basic Energy Sciences for support under grant No. DE-SC0014407. 

\bibliography{all_cited_bib}

\begin{thebibliography}{78}%
\makeatletter
\providecommand \@ifxundefined [1]{%
 \@ifx{#1\undefined}
}%
\providecommand \@ifnum [1]{%
 \ifnum #1\expandafter \@firstoftwo
 \else \expandafter \@secondoftwo
 \fi
}%
\providecommand \@ifx [1]{%
 \ifx #1\expandafter \@firstoftwo
 \else \expandafter \@secondoftwo
 \fi
}%
\providecommand \natexlab [1]{#1}%
\providecommand \enquote  [1]{``#1''}%
\providecommand \bibnamefont  [1]{#1}%
\providecommand \bibfnamefont [1]{#1}%
\providecommand \citenamefont [1]{#1}%
\providecommand \href@noop [0]{\@secondoftwo}%
\providecommand \href [0]{\begingroup \@sanitize@url \@href}%
\providecommand \@href[1]{\@@startlink{#1}\@@href}%
\providecommand \@@href[1]{\endgroup#1\@@endlink}%
\providecommand \@sanitize@url [0]{\catcode `\\12\catcode `\$12\catcode
  `\&12\catcode `\#12\catcode `\^12\catcode `\_12\catcode `\%12\relax}%
\providecommand \@@startlink[1]{}%
\providecommand \@@endlink[0]{}%
\providecommand \url  [0]{\begingroup\@sanitize@url \@url }%
\providecommand \@url [1]{\endgroup\@href {#1}{\urlprefix }}%
\providecommand \urlprefix  [0]{URL }%
\providecommand \Eprint [0]{\href }%
\providecommand \doibase [0]{http://dx.doi.org/}%
\providecommand \selectlanguage [0]{\@gobble}%
\providecommand \bibinfo  [0]{\@secondoftwo}%
\providecommand \bibfield  [0]{\@secondoftwo}%
\providecommand \translation [1]{[#1]}%
\providecommand \BibitemOpen [0]{}%
\providecommand \bibitemStop [0]{}%
\providecommand \bibitemNoStop [0]{.\EOS\space}%
\providecommand \EOS [0]{\spacefactor3000\relax}%
\providecommand \BibitemShut  [1]{\csname bibitem#1\endcsname}%
\let\auto@bib@innerbib\@empty
\bibitem [{\citenamefont {Cavalieri}\ \emph {et~al.}(2007)\citenamefont
  {Cavalieri}, \citenamefont {M\"{u}ller}, \citenamefont {Uphues},
  \citenamefont {Yakovlev}, \citenamefont {Baltuska}, \citenamefont {Horvath},
  \citenamefont {Schmidt}, \citenamefont {Bl\"{u}mel}, \citenamefont
  {Holzwarth}, \citenamefont {Hendel}, \citenamefont {Drescher}, \citenamefont
  {Kleineberg}, \citenamefont {Echenique}, \citenamefont {Kienberger},
  \citenamefont {Krausz}, \citenamefont {Heinzmann}, \citenamefont {Blu},
  \citenamefont {Uphues}, \citenamefont {Yakovlev}, \citenamefont {Baltus},\
  and\ \citenamefont {Heinzmann}}]{Cavalieri2007}%
  \BibitemOpen
  \bibfield  {author} {\bibinfo {author} {\bibfnamefont {A.~L.}\ \bibnamefont
  {Cavalieri}}, \bibinfo {author} {\bibfnamefont {N.}~\bibnamefont
  {M\"{u}ller}}, \bibinfo {author} {\bibfnamefont {T.}~\bibnamefont {Uphues}},
  \bibinfo {author} {\bibfnamefont {V.~S.}\ \bibnamefont {Yakovlev}}, \bibinfo
  {author} {\bibfnamefont {A.}~\bibnamefont {Baltuska}}, \bibinfo {author}
  {\bibfnamefont {B.}~\bibnamefont {Horvath}}, \bibinfo {author} {\bibfnamefont
  {B.}~\bibnamefont {Schmidt}}, \bibinfo {author} {\bibfnamefont
  {L.}~\bibnamefont {Bl\"{u}mel}}, \bibinfo {author} {\bibfnamefont
  {R.}~\bibnamefont {Holzwarth}}, \bibinfo {author} {\bibfnamefont
  {S.}~\bibnamefont {Hendel}}, \bibinfo {author} {\bibfnamefont
  {M.}~\bibnamefont {Drescher}}, \bibinfo {author} {\bibfnamefont
  {U.}~\bibnamefont {Kleineberg}}, \bibinfo {author} {\bibfnamefont {P.~M.}\
  \bibnamefont {Echenique}}, \bibinfo {author} {\bibfnamefont {R.}~\bibnamefont
  {Kienberger}}, \bibinfo {author} {\bibfnamefont {F.}~\bibnamefont {Krausz}},
  \bibinfo {author} {\bibfnamefont {U.}~\bibnamefont {Heinzmann}}, \bibinfo
  {author} {\bibfnamefont {L.}~\bibnamefont {Blu}}, \bibinfo {author}
  {\bibfnamefont {T.}~\bibnamefont {Uphues}}, \bibinfo {author} {\bibfnamefont
  {V.~S.}\ \bibnamefont {Yakovlev}}, \bibinfo {author} {\bibfnamefont
  {A.}~\bibnamefont {Baltus}}, \ and\ \bibinfo {author} {\bibfnamefont
  {U.}~\bibnamefont {Heinzmann}},\ }\href {\doibase 10.1038/nature06229}
  {\bibfield  {journal} {\bibinfo  {journal} {Nature}\ }\textbf {\bibinfo
  {volume} {449}},\ \bibinfo {pages} {1029} (\bibinfo {year}
  {2007})}\BibitemShut {NoStop}%
\bibitem [{\citenamefont {Orenstein}(2012)}]{Orenstein2012c}%
  \BibitemOpen
  \bibfield  {author} {\bibinfo {author} {\bibfnamefont {J.}~\bibnamefont
  {Orenstein}},\ }\href {\doibase 10.1063/PT.3.1717} {\bibfield  {journal}
  {\bibinfo  {journal} {Physics Today}\ }\textbf {\bibinfo {volume} {65}},\
  \bibinfo {pages} {44} (\bibinfo {year} {2012})}\BibitemShut {NoStop}%
\bibitem [{\citenamefont {Chollet}\ \emph {et~al.}(2005)\citenamefont
  {Chollet}, \citenamefont {Guerin}, \citenamefont {Uchida}, \citenamefont
  {Fukaya}, \citenamefont {Shimoda}, \citenamefont {Ishikawa}, \citenamefont
  {Matsuda}, \citenamefont {Hasegawa}, \citenamefont {Ota}, \citenamefont
  {Yamochi}, \citenamefont {Saito}, \citenamefont {Tazaki}, \citenamefont
  {Adachi},\ and\ \citenamefont {Koshihara}}]{Chollet2005}%
  \BibitemOpen
  \bibfield  {author} {\bibinfo {author} {\bibfnamefont {M.}~\bibnamefont
  {Chollet}}, \bibinfo {author} {\bibfnamefont {L.}~\bibnamefont {Guerin}},
  \bibinfo {author} {\bibfnamefont {N.}~\bibnamefont {Uchida}}, \bibinfo
  {author} {\bibfnamefont {S.}~\bibnamefont {Fukaya}}, \bibinfo {author}
  {\bibfnamefont {H.}~\bibnamefont {Shimoda}}, \bibinfo {author} {\bibfnamefont
  {T.}~\bibnamefont {Ishikawa}}, \bibinfo {author} {\bibfnamefont
  {K.}~\bibnamefont {Matsuda}}, \bibinfo {author} {\bibfnamefont
  {T.}~\bibnamefont {Hasegawa}}, \bibinfo {author} {\bibfnamefont
  {A.}~\bibnamefont {Ota}}, \bibinfo {author} {\bibfnamefont {H.}~\bibnamefont
  {Yamochi}}, \bibinfo {author} {\bibfnamefont {G.}~\bibnamefont {Saito}},
  \bibinfo {author} {\bibfnamefont {R.}~\bibnamefont {Tazaki}}, \bibinfo
  {author} {\bibfnamefont {S.-I.}\ \bibnamefont {Adachi}}, \ and\ \bibinfo
  {author} {\bibfnamefont {S.-Y.}\ \bibnamefont {Koshihara}},\ }\href {\doibase
  10.1126/science.1105067} {\bibfield  {journal} {\bibinfo  {journal} {Science
  (New York, N.Y.)}\ }\textbf {\bibinfo {volume} {307}},\ \bibinfo {pages} {86}
  (\bibinfo {year} {2005})}\BibitemShut {NoStop}%
\bibitem [{\citenamefont {Corkum}\ and\ \citenamefont
  {Krausz}(2007)}]{Corkum2007a}%
  \BibitemOpen
  \bibfield  {author} {\bibinfo {author} {\bibfnamefont {P.}~\bibnamefont
  {Corkum}}\ and\ \bibinfo {author} {\bibfnamefont {F.}~\bibnamefont
  {Krausz}},\ }\href
  {http://www.nature.com/nphys/journal/v3/n6/abs/nphys620.html} {\bibfield
  {journal} {\bibinfo  {journal} {Nature Physics}\ }\textbf {\bibinfo {volume}
  {3}},\ \bibinfo {pages} {381} (\bibinfo {year} {2007})}\BibitemShut {NoStop}%
\bibitem [{\citenamefont {Onda}\ \emph {et~al.}(2008)\citenamefont {Onda},
  \citenamefont {Ogihara}, \citenamefont {Yonemitsu}, \citenamefont {Maeshima},
  \citenamefont {Ishikawa}, \citenamefont {Okimoto}, \citenamefont {Shao},
  \citenamefont {Nakano}, \citenamefont {Yamochi}, \citenamefont {Saito},\ and\
  \citenamefont {Koshihara}}]{Onda2008}%
  \BibitemOpen
  \bibfield  {author} {\bibinfo {author} {\bibfnamefont {K.}~\bibnamefont
  {Onda}}, \bibinfo {author} {\bibfnamefont {S.}~\bibnamefont {Ogihara}},
  \bibinfo {author} {\bibfnamefont {K.}~\bibnamefont {Yonemitsu}}, \bibinfo
  {author} {\bibfnamefont {N.}~\bibnamefont {Maeshima}}, \bibinfo {author}
  {\bibfnamefont {T.}~\bibnamefont {Ishikawa}}, \bibinfo {author}
  {\bibfnamefont {Y.}~\bibnamefont {Okimoto}}, \bibinfo {author} {\bibfnamefont
  {X.}~\bibnamefont {Shao}}, \bibinfo {author} {\bibfnamefont {Y.}~\bibnamefont
  {Nakano}}, \bibinfo {author} {\bibfnamefont {H.}~\bibnamefont {Yamochi}},
  \bibinfo {author} {\bibfnamefont {G.}~\bibnamefont {Saito}}, \ and\ \bibinfo
  {author} {\bibfnamefont {S.-y.}\ \bibnamefont {Koshihara}},\ }\href {\doibase
  10.1103/PhysRevLett.101.067403} {\bibfield  {journal} {\bibinfo  {journal}
  {Phys. Rev. Lett.}\ }\textbf {\bibinfo {volume} {101}},\ \bibinfo {pages}
  {067403} (\bibinfo {year} {2008})}\BibitemShut {NoStop}%
\bibitem [{\citenamefont {Kampfrath}\ \emph {et~al.}(2013)\citenamefont
  {Kampfrath}, \citenamefont {Tanaka},\ and\ \citenamefont
  {Nelson}}]{Kampfrath2013}%
  \BibitemOpen
  \bibfield  {author} {\bibinfo {author} {\bibfnamefont {T.}~\bibnamefont
  {Kampfrath}}, \bibinfo {author} {\bibfnamefont {K.}~\bibnamefont {Tanaka}}, \
  and\ \bibinfo {author} {\bibfnamefont {K.~A.}\ \bibnamefont {Nelson}},\
  }\href {http://dx.doi.org/10.1038/nphoton.2013.184} {\bibfield  {journal}
  {\bibinfo  {journal} {Nature Photonics}\ }\textbf {\bibinfo {volume} {7}},\
  \bibinfo {pages} {680 EP } (\bibinfo {year} {2013})},\ \bibinfo {note}
  {review Article L3 -}\BibitemShut {NoStop}%
\bibitem [{\citenamefont {Fausti}\ \emph {et~al.}(2011)\citenamefont {Fausti},
  \citenamefont {Tobey}, \citenamefont {Dean}, \citenamefont {Kaiser},
  \citenamefont {Dienst}, \citenamefont {Hoffmann}, \citenamefont {Pyon},
  \citenamefont {Takayama}, \citenamefont {Takagi},\ and\ \citenamefont
  {Cavalleri}}]{Fausti2011}%
  \BibitemOpen
  \bibfield  {author} {\bibinfo {author} {\bibfnamefont {D.}~\bibnamefont
  {Fausti}}, \bibinfo {author} {\bibfnamefont {R.~I.}\ \bibnamefont {Tobey}},
  \bibinfo {author} {\bibfnamefont {N.}~\bibnamefont {Dean}}, \bibinfo {author}
  {\bibfnamefont {S.}~\bibnamefont {Kaiser}}, \bibinfo {author} {\bibfnamefont
  {A.}~\bibnamefont {Dienst}}, \bibinfo {author} {\bibfnamefont {M.~C.}\
  \bibnamefont {Hoffmann}}, \bibinfo {author} {\bibfnamefont {S.}~\bibnamefont
  {Pyon}}, \bibinfo {author} {\bibfnamefont {T.}~\bibnamefont {Takayama}},
  \bibinfo {author} {\bibfnamefont {H.}~\bibnamefont {Takagi}}, \ and\ \bibinfo
  {author} {\bibfnamefont {A.}~\bibnamefont {Cavalleri}},\ }\href {\doibase
  10.1126/science.1197294} {\bibfield  {journal} {\bibinfo  {journal}
  {Science}\ }\textbf {\bibinfo {volume} {331}},\ \bibinfo {pages} {189}
  (\bibinfo {year} {2011})}\BibitemShut {NoStop}%
\bibitem [{\citenamefont {Polli}\ \emph {et~al.}(2007)\citenamefont {Polli},
  \citenamefont {Rini}, \citenamefont {Wall}, \citenamefont {Schoenlein},
  \citenamefont {Tomioka}, \citenamefont {Tokura}, \citenamefont {Cerullo},\
  and\ \citenamefont {Cavalleri}}]{Polli2007}%
  \BibitemOpen
  \bibfield  {author} {\bibinfo {author} {\bibfnamefont {D.}~\bibnamefont
  {Polli}}, \bibinfo {author} {\bibfnamefont {M.}~\bibnamefont {Rini}},
  \bibinfo {author} {\bibfnamefont {S.}~\bibnamefont {Wall}}, \bibinfo {author}
  {\bibfnamefont {R.~W.}\ \bibnamefont {Schoenlein}}, \bibinfo {author}
  {\bibfnamefont {Y.}~\bibnamefont {Tomioka}}, \bibinfo {author} {\bibfnamefont
  {Y.}~\bibnamefont {Tokura}}, \bibinfo {author} {\bibfnamefont
  {G.}~\bibnamefont {Cerullo}}, \ and\ \bibinfo {author} {\bibfnamefont
  {A.}~\bibnamefont {Cavalleri}},\ }\href {http://dx.doi.org/10.1038/nmat1979}
  {\bibfield  {journal} {\bibinfo  {journal} {Nature Materials}\ }\textbf
  {\bibinfo {volume} {6}},\ \bibinfo {pages} {643 EP } (\bibinfo {year}
  {2007})}\BibitemShut {NoStop}%
\bibitem [{\citenamefont {Ehrke}\ \emph {et~al.}(2011)\citenamefont {Ehrke},
  \citenamefont {Tobey}, \citenamefont {Wall}, \citenamefont {Cavill},
  \citenamefont {F\"orst}, \citenamefont {Khanna}, \citenamefont {Garl},
  \citenamefont {Stojanovic}, \citenamefont {Prabhakaran}, \citenamefont
  {Boothroyd}, \citenamefont {Gensch}, \citenamefont {Mirone}, \citenamefont
  {Reutler}, \citenamefont {Revcolevschi}, \citenamefont {Dhesi},\ and\
  \citenamefont {Cavalleri}}]{Ehrke2011}%
  \BibitemOpen
  \bibfield  {author} {\bibinfo {author} {\bibfnamefont {H.}~\bibnamefont
  {Ehrke}}, \bibinfo {author} {\bibfnamefont {R.~I.}\ \bibnamefont {Tobey}},
  \bibinfo {author} {\bibfnamefont {S.}~\bibnamefont {Wall}}, \bibinfo {author}
  {\bibfnamefont {S.~A.}\ \bibnamefont {Cavill}}, \bibinfo {author}
  {\bibfnamefont {M.}~\bibnamefont {F\"orst}}, \bibinfo {author} {\bibfnamefont
  {V.}~\bibnamefont {Khanna}}, \bibinfo {author} {\bibfnamefont
  {T.}~\bibnamefont {Garl}}, \bibinfo {author} {\bibfnamefont {N.}~\bibnamefont
  {Stojanovic}}, \bibinfo {author} {\bibfnamefont {D.}~\bibnamefont
  {Prabhakaran}}, \bibinfo {author} {\bibfnamefont {A.~T.}\ \bibnamefont
  {Boothroyd}}, \bibinfo {author} {\bibfnamefont {M.}~\bibnamefont {Gensch}},
  \bibinfo {author} {\bibfnamefont {A.}~\bibnamefont {Mirone}}, \bibinfo
  {author} {\bibfnamefont {P.}~\bibnamefont {Reutler}}, \bibinfo {author}
  {\bibfnamefont {A.}~\bibnamefont {Revcolevschi}}, \bibinfo {author}
  {\bibfnamefont {S.~S.}\ \bibnamefont {Dhesi}}, \ and\ \bibinfo {author}
  {\bibfnamefont {A.}~\bibnamefont {Cavalleri}},\ }\href {\doibase
  10.1103/PhysRevLett.106.217401} {\bibfield  {journal} {\bibinfo  {journal}
  {Phys. Rev. Lett.}\ }\textbf {\bibinfo {volume} {106}},\ \bibinfo {pages}
  {217401} (\bibinfo {year} {2011})}\BibitemShut {NoStop}%
\bibitem [{\citenamefont {Zhang}\ \emph {et~al.}(2016)\citenamefont {Zhang},
  \citenamefont {Tan}, \citenamefont {Liu}, \citenamefont {Teitelbaum},
  \citenamefont {Post}, \citenamefont {Jin}, \citenamefont {Nelson},
  \citenamefont {Basov}, \citenamefont {Wu},\ and\ \citenamefont
  {Averitt}}]{Zhang2016}%
  \BibitemOpen
  \bibfield  {author} {\bibinfo {author} {\bibfnamefont {J.}~\bibnamefont
  {Zhang}}, \bibinfo {author} {\bibfnamefont {X.}~\bibnamefont {Tan}}, \bibinfo
  {author} {\bibfnamefont {M.}~\bibnamefont {Liu}}, \bibinfo {author}
  {\bibfnamefont {S.~W.}\ \bibnamefont {Teitelbaum}}, \bibinfo {author}
  {\bibfnamefont {K.~W.}\ \bibnamefont {Post}}, \bibinfo {author}
  {\bibfnamefont {F.}~\bibnamefont {Jin}}, \bibinfo {author} {\bibfnamefont
  {K.~.~A.}\ \bibnamefont {Nelson}}, \bibinfo {author} {\bibfnamefont {D.~N.}\
  \bibnamefont {Basov}}, \bibinfo {author} {\bibfnamefont {W.}~\bibnamefont
  {Wu}}, \ and\ \bibinfo {author} {\bibfnamefont {R.~D.}\ \bibnamefont
  {Averitt}},\ }\href {http://dx.doi.org/10.1038/nmat4695} {\bibfield
  {journal} {\bibinfo  {journal} {Nature Materials}\ }\textbf {\bibinfo
  {volume} {15}},\ \bibinfo {pages} {956 EP } (\bibinfo {year}
  {2016})}\BibitemShut {NoStop}%
\bibitem [{\citenamefont {Casals}\ \emph {et~al.}(2016)\citenamefont {Casals},
  \citenamefont {Cichelero}, \citenamefont {Garc\'{\i}a~Fern\'andez},
  \citenamefont {Junquera}, \citenamefont {Pesquera}, \citenamefont
  {Campoy-Quiles}, \citenamefont {Infante}, \citenamefont {S\'anchez},
  \citenamefont {Fontcuberta},\ and\ \citenamefont {Herranz}}]{Casals2016}%
  \BibitemOpen
  \bibfield  {author} {\bibinfo {author} {\bibfnamefont {B.}~\bibnamefont
  {Casals}}, \bibinfo {author} {\bibfnamefont {R.}~\bibnamefont {Cichelero}},
  \bibinfo {author} {\bibfnamefont {P.}~\bibnamefont
  {Garc\'{\i}a~Fern\'andez}}, \bibinfo {author} {\bibfnamefont
  {J.}~\bibnamefont {Junquera}}, \bibinfo {author} {\bibfnamefont
  {D.}~\bibnamefont {Pesquera}}, \bibinfo {author} {\bibfnamefont
  {M.}~\bibnamefont {Campoy-Quiles}}, \bibinfo {author} {\bibfnamefont {I.~C.}\
  \bibnamefont {Infante}}, \bibinfo {author} {\bibfnamefont {F.}~\bibnamefont
  {S\'anchez}}, \bibinfo {author} {\bibfnamefont {J.}~\bibnamefont
  {Fontcuberta}}, \ and\ \bibinfo {author} {\bibfnamefont {G.}~\bibnamefont
  {Herranz}},\ }\href {\doibase 10.1103/PhysRevLett.117.026401} {\bibfield
  {journal} {\bibinfo  {journal} {Phys. Rev. Lett.}\ }\textbf {\bibinfo
  {volume} {117}},\ \bibinfo {pages} {026401} (\bibinfo {year}
  {2016})}\BibitemShut {NoStop}%
\bibitem [{\citenamefont {Okamoto}\ \emph {et~al.}(2007)\citenamefont
  {Okamoto}, \citenamefont {Matsuzaki}, \citenamefont {Wakabayashi},
  \citenamefont {Takahashi},\ and\ \citenamefont {Hasegawa}}]{Okamoto2007b}%
  \BibitemOpen
  \bibfield  {author} {\bibinfo {author} {\bibfnamefont {H.}~\bibnamefont
  {Okamoto}}, \bibinfo {author} {\bibfnamefont {H.}~\bibnamefont {Matsuzaki}},
  \bibinfo {author} {\bibfnamefont {T.}~\bibnamefont {Wakabayashi}}, \bibinfo
  {author} {\bibfnamefont {Y.}~\bibnamefont {Takahashi}}, \ and\ \bibinfo
  {author} {\bibfnamefont {T.}~\bibnamefont {Hasegawa}},\ }\href {\doibase
  10.1103/PhysRevLett.98.037401} {\bibfield  {journal} {\bibinfo  {journal}
  {Phys. Rev. Lett.}\ }\textbf {\bibinfo {volume} {98}},\ \bibinfo {pages}
  {037401} (\bibinfo {year} {2007})}\BibitemShut {NoStop}%
\bibitem [{\citenamefont {Plummer}(1997)}]{Plummer1997}%
  \BibitemOpen
  \bibfield  {author} {\bibinfo {author} {\bibfnamefont {W.}~\bibnamefont
  {Plummer}},\ }\href {\doibase 10.1126/science.277.5331.1447} {\bibfield
  {journal} {\bibinfo  {journal} {Science}\ }\textbf {\bibinfo {volume}
  {277}},\ \bibinfo {pages} {1447} (\bibinfo {year} {1997})}\BibitemShut
  {NoStop}%
\bibitem [{\citenamefont {Ramakrishna}\ \emph {et~al.}(2001)\citenamefont
  {Ramakrishna}, \citenamefont {Willig},\ and\ \citenamefont
  {May}}]{ramakrishna2001}%
  \BibitemOpen
  \bibfield  {author} {\bibinfo {author} {\bibfnamefont {S.}~\bibnamefont
  {Ramakrishna}}, \bibinfo {author} {\bibfnamefont {F.}~\bibnamefont {Willig}},
  \ and\ \bibinfo {author} {\bibfnamefont {V.}~\bibnamefont {May}},\ }\href
  {\doibase https://doi.org/10.1063/1.1386433} {\bibfield  {journal} {\bibinfo
  {journal} {J. Chem. Phys.}\ }\textbf {\bibinfo {volume} {115}},\ \bibinfo
  {pages} {2743} (\bibinfo {year} {2001})}\BibitemShut {NoStop}%
\bibitem [{\citenamefont {Domcke}(1991)}]{domcke1991}%
  \BibitemOpen
  \bibfield  {author} {\bibinfo {author} {\bibfnamefont {W.}~\bibnamefont
  {Domcke}},\ }\href {\doibase https://doi.org/10.1016/0370-1573(91)90125-6}
  {\bibfield  {journal} {\bibinfo  {journal} {Phys. Rep.}\ }\textbf {\bibinfo
  {volume} {208}},\ \bibinfo {pages} {97} (\bibinfo {year} {1991})}\BibitemShut
  {NoStop}%
\bibitem [{\citenamefont {Sebastian}\ and\ \citenamefont
  {Tachiya}(2006)}]{sebastian2006}%
  \BibitemOpen
  \bibfield  {author} {\bibinfo {author} {\bibfnamefont {K.~L.}\ \bibnamefont
  {Sebastian}}\ and\ \bibinfo {author} {\bibfnamefont {M.}~\bibnamefont
  {Tachiya}},\ }\href {\doibase https://doi.org/10.1063/1.2171238} {\bibfield
  {journal} {\bibinfo  {journal} {J. Chem. Phys.}\ }\textbf {\bibinfo {volume}
  {124}},\ \bibinfo {pages} {064713} (\bibinfo {year} {2006})}\BibitemShut
  {NoStop}%
\bibitem [{\citenamefont {Smallwood}\ \emph {et~al.}(2016)\citenamefont
  {Smallwood}, \citenamefont {Kaindl},\ and\ \citenamefont
  {Lanzara}}]{Smallwood2016}%
  \BibitemOpen
  \bibfield  {author} {\bibinfo {author} {\bibfnamefont {C.~L.}\ \bibnamefont
  {Smallwood}}, \bibinfo {author} {\bibfnamefont {R.~A.}\ \bibnamefont
  {Kaindl}}, \ and\ \bibinfo {author} {\bibfnamefont {A.}~\bibnamefont
  {Lanzara}},\ }\href {http://stacks.iop.org/0295-5075/115/i=2/a=27001}
  {\bibfield  {journal} {\bibinfo  {journal} {EPL (Europhysics Letters)}\
  }\textbf {\bibinfo {volume} {115}},\ \bibinfo {pages} {27001} (\bibinfo
  {year} {2016})}\BibitemShut {NoStop}%
\bibitem [{\citenamefont {Freericks}\ \emph {et~al.}(2009)\citenamefont
  {Freericks}, \citenamefont {Krishnamurthy},\ and\ \citenamefont
  {Pruschke}}]{Freericks2009}%
  \BibitemOpen
  \bibfield  {author} {\bibinfo {author} {\bibfnamefont {J.~K.}\ \bibnamefont
  {Freericks}}, \bibinfo {author} {\bibfnamefont {H.~R.}\ \bibnamefont
  {Krishnamurthy}}, \ and\ \bibinfo {author} {\bibfnamefont {T.}~\bibnamefont
  {Pruschke}},\ }\href {\doibase 10.1103/PhysRevLett.102.136401} {\bibfield
  {journal} {\bibinfo  {journal} {Phys. Rev. Lett.}\ }\textbf {\bibinfo
  {volume} {102}},\ \bibinfo {pages} {136401} (\bibinfo {year}
  {2009})}\BibitemShut {NoStop}%
\bibitem [{\citenamefont {Shao}\ \emph {et~al.}(2016)\citenamefont {Shao},
  \citenamefont {Tohyama}, \citenamefont {Luo},\ and\ \citenamefont
  {Lu}}]{Shao2016}%
  \BibitemOpen
  \bibfield  {author} {\bibinfo {author} {\bibfnamefont {C.}~\bibnamefont
  {Shao}}, \bibinfo {author} {\bibfnamefont {T.}~\bibnamefont {Tohyama}},
  \bibinfo {author} {\bibfnamefont {H.-G.}\ \bibnamefont {Luo}}, \ and\
  \bibinfo {author} {\bibfnamefont {H.}~\bibnamefont {Lu}},\ }\href {\doibase
  10.1103/PhysRevB.93.195144} {\bibfield  {journal} {\bibinfo  {journal} {Phys.
  Rev. B}\ }\textbf {\bibinfo {volume} {93}},\ \bibinfo {pages} {195144}
  (\bibinfo {year} {2016})}\BibitemShut {NoStop}%
\bibitem [{\citenamefont {Carpentier}\ \emph {et~al.}(2002)\citenamefont
  {Carpentier}, \citenamefont {Pe\ifmmode~\mbox{\c{c}}\else \c{c}\fi{}a},\ and\
  \citenamefont {Balents}}]{Carpentier2002}%
  \BibitemOpen
  \bibfield  {author} {\bibinfo {author} {\bibfnamefont {D.}~\bibnamefont
  {Carpentier}}, \bibinfo {author} {\bibfnamefont {C.}~\bibnamefont
  {Pe\ifmmode~\mbox{\c{c}}\else \c{c}\fi{}a}}, \ and\ \bibinfo {author}
  {\bibfnamefont {L.}~\bibnamefont {Balents}},\ }\href {\doibase
  10.1103/PhysRevB.66.153304} {\bibfield  {journal} {\bibinfo  {journal} {Phys.
  Rev. B}\ }\textbf {\bibinfo {volume} {66}},\ \bibinfo {pages} {153304}
  (\bibinfo {year} {2002})}\BibitemShut {NoStop}%
\bibitem [{\citenamefont {Auslaender}\ \emph {et~al.}(2002)\citenamefont
  {Auslaender}, \citenamefont {Yacoby}, \citenamefont {de~Picciotto},
  \citenamefont {Baldwin}, \citenamefont {Pfeiffer},\ and\ \citenamefont
  {West}}]{Auslaender2002}%
  \BibitemOpen
  \bibfield  {author} {\bibinfo {author} {\bibfnamefont {O.~M.}\ \bibnamefont
  {Auslaender}}, \bibinfo {author} {\bibfnamefont {A.}~\bibnamefont {Yacoby}},
  \bibinfo {author} {\bibfnamefont {R.}~\bibnamefont {de~Picciotto}}, \bibinfo
  {author} {\bibfnamefont {K.~W.}\ \bibnamefont {Baldwin}}, \bibinfo {author}
  {\bibfnamefont {L.~N.}\ \bibnamefont {Pfeiffer}}, \ and\ \bibinfo {author}
  {\bibfnamefont {K.~W.}\ \bibnamefont {West}},\ }\href {\doibase
  10.1126/science.1066266} {\bibfield  {journal} {\bibinfo  {journal}
  {Science}\ }\textbf {\bibinfo {volume} {295}},\ \bibinfo {pages} {825}
  (\bibinfo {year} {2002})}\BibitemShut {NoStop}%
\bibitem [{\citenamefont {Auslaender}\ \emph {et~al.}(2005)\citenamefont
  {Auslaender}, \citenamefont {Steinberg}, \citenamefont {Yacoby},
  \citenamefont {Tserkovnyak}, \citenamefont {Halperin}, \citenamefont
  {Baldwin}, \citenamefont {Pfeiffer},\ and\ \citenamefont
  {West}}]{Auslaender2005}%
  \BibitemOpen
  \bibfield  {author} {\bibinfo {author} {\bibfnamefont {O.~M.}\ \bibnamefont
  {Auslaender}}, \bibinfo {author} {\bibfnamefont {H.}~\bibnamefont
  {Steinberg}}, \bibinfo {author} {\bibfnamefont {A.}~\bibnamefont {Yacoby}},
  \bibinfo {author} {\bibfnamefont {Y.}~\bibnamefont {Tserkovnyak}}, \bibinfo
  {author} {\bibfnamefont {B.~I.}\ \bibnamefont {Halperin}}, \bibinfo {author}
  {\bibfnamefont {K.~W.}\ \bibnamefont {Baldwin}}, \bibinfo {author}
  {\bibfnamefont {L.~N.}\ \bibnamefont {Pfeiffer}}, \ and\ \bibinfo {author}
  {\bibfnamefont {K.~W.}\ \bibnamefont {West}},\ }\href {\doibase
  10.1126/science.1107821} {\bibfield  {journal} {\bibinfo  {journal}
  {Science}\ }\textbf {\bibinfo {volume} {308}},\ \bibinfo {pages} {88}
  (\bibinfo {year} {2005})}\BibitemShut {NoStop}%
\bibitem [{\citenamefont {Cohen}\ \emph {et~al.}(2014)\citenamefont {Cohen},
  \citenamefont {Gull}, \citenamefont {Reichman},\ and\ \citenamefont
  {Millis}}]{Cohen2014}%
  \BibitemOpen
  \bibfield  {author} {\bibinfo {author} {\bibfnamefont {G.}~\bibnamefont
  {Cohen}}, \bibinfo {author} {\bibfnamefont {E.}~\bibnamefont {Gull}},
  \bibinfo {author} {\bibfnamefont {D.~R.}\ \bibnamefont {Reichman}}, \ and\
  \bibinfo {author} {\bibfnamefont {A.~J.}\ \bibnamefont {Millis}},\ }\href
  {\doibase 10.1103/PhysRevLett.112.146802} {\bibfield  {journal} {\bibinfo
  {journal} {Phys. Rev. Lett.}\ }\textbf {\bibinfo {volume} {112}},\ \bibinfo
  {pages} {146802} (\bibinfo {year} {2014})}\BibitemShut {NoStop}%
\bibitem [{\citenamefont {Kantian}\ \emph {et~al.}(2015)\citenamefont
  {Kantian}, \citenamefont {Schollw\"ock},\ and\ \citenamefont
  {Giamarchi}}]{Kantian2015}%
  \BibitemOpen
  \bibfield  {author} {\bibinfo {author} {\bibfnamefont {A.}~\bibnamefont
  {Kantian}}, \bibinfo {author} {\bibfnamefont {U.}~\bibnamefont
  {Schollw\"ock}}, \ and\ \bibinfo {author} {\bibfnamefont {T.}~\bibnamefont
  {Giamarchi}},\ }\href {\doibase 10.1103/PhysRevLett.115.165301} {\bibfield
  {journal} {\bibinfo  {journal} {Phys. Rev. Lett.}\ }\textbf {\bibinfo
  {volume} {115}},\ \bibinfo {pages} {165301} (\bibinfo {year}
  {2015})}\BibitemShut {NoStop}%
\bibitem [{\citenamefont {Bohrdt}\ \emph {et~al.}(2018)\citenamefont {Bohrdt},
  \citenamefont {Greif}, \citenamefont {Demler}, \citenamefont {Knap},\ and\
  \citenamefont {Grusdt}}]{Bohrdt2017}%
  \BibitemOpen
  \bibfield  {author} {\bibinfo {author} {\bibfnamefont {A.}~\bibnamefont
  {Bohrdt}}, \bibinfo {author} {\bibfnamefont {D.}~\bibnamefont {Greif}},
  \bibinfo {author} {\bibfnamefont {E.}~\bibnamefont {Demler}}, \bibinfo
  {author} {\bibfnamefont {M.}~\bibnamefont {Knap}}, \ and\ \bibinfo {author}
  {\bibfnamefont {F.}~\bibnamefont {Grusdt}},\ }\href {\doibase
  10.1103/PhysRevB.97.125117} {\bibfield  {journal} {\bibinfo  {journal} {Phys.
  Rev. B}\ }\textbf {\bibinfo {volume} {97}},\ \bibinfo {pages} {125117}
  (\bibinfo {year} {2018})}\BibitemShut {NoStop}%
\bibitem [{\citenamefont {Brown}\ \emph {et~al.}(1995)\citenamefont {Brown},
  \citenamefont {Pomp}, \citenamefont {Hart},\ and\ \citenamefont
  {de~Leeuw}}]{brown1995}%
  \BibitemOpen
  \bibfield  {author} {\bibinfo {author} {\bibfnamefont {A.~R.}\ \bibnamefont
  {Brown}}, \bibinfo {author} {\bibfnamefont {A.}~\bibnamefont {Pomp}},
  \bibinfo {author} {\bibfnamefont {C.~M.}\ \bibnamefont {Hart}}, \ and\
  \bibinfo {author} {\bibfnamefont {D.~M.}\ \bibnamefont {de~Leeuw}},\ }\href
  {\doibase 10.1126/science.270.5238.972} {\bibfield  {journal} {\bibinfo
  {journal} {Science}\ }\textbf {\bibinfo {volume} {270}},\ \bibinfo {pages}
  {972} (\bibinfo {year} {1995})}\BibitemShut {NoStop}%
\bibitem [{\citenamefont {Burroughes}\ \emph {et~al.}(1990)\citenamefont
  {Burroughes}, \citenamefont {Bradeley}, \citenamefont {Brown}, \citenamefont
  {Marks}, \citenamefont {Machey}, \citenamefont {Friend}, \citenamefont
  {Burns},\ and\ \citenamefont {Holmes}}]{burroughes1990}%
  \BibitemOpen
  \bibfield  {author} {\bibinfo {author} {\bibfnamefont {J.~H.}\ \bibnamefont
  {Burroughes}}, \bibinfo {author} {\bibfnamefont {D.~D.~C.}\ \bibnamefont
  {Bradeley}}, \bibinfo {author} {\bibfnamefont {A.~R.}\ \bibnamefont {Brown}},
  \bibinfo {author} {\bibfnamefont {R.~N.}\ \bibnamefont {Marks}}, \bibinfo
  {author} {\bibfnamefont {K.}~\bibnamefont {Machey}}, \bibinfo {author}
  {\bibfnamefont {R.~H.}\ \bibnamefont {Friend}}, \bibinfo {author}
  {\bibfnamefont {P.~L.}\ \bibnamefont {Burns}}, \ and\ \bibinfo {author}
  {\bibfnamefont {A.~B.}\ \bibnamefont {Holmes}},\ }\href {\doibase
  https://doi.org/10.1038/347539a0} {\bibfield  {journal} {\bibinfo  {journal}
  {Nature}\ }\textbf {\bibinfo {volume} {347}},\ \bibinfo {pages} {539}
  (\bibinfo {year} {1990})}\BibitemShut {NoStop}%
\bibitem [{\citenamefont {Dodabalapur}\ \emph
  {et~al.}(1995{\natexlab{a}})\citenamefont {Dodabalapur}, \citenamefont
  {Torsi},\ and\ \citenamefont {Katz}}]{dodabalapur1995}%
  \BibitemOpen
  \bibfield  {author} {\bibinfo {author} {\bibfnamefont {A.}~\bibnamefont
  {Dodabalapur}}, \bibinfo {author} {\bibfnamefont {L.}~\bibnamefont {Torsi}},
  \ and\ \bibinfo {author} {\bibfnamefont {H.~E.}\ \bibnamefont {Katz}},\
  }\href {\doibase 10.1126/science.268.5208.270} {\bibfield  {journal}
  {\bibinfo  {journal} {Science}\ }\textbf {\bibinfo {volume} {268}},\ \bibinfo
  {pages} {270} (\bibinfo {year} {1995}{\natexlab{a}})}\BibitemShut {NoStop}%
\bibitem [{\citenamefont {Dodabalapur}\ \emph
  {et~al.}(1995{\natexlab{b}})\citenamefont {Dodabalapur}, \citenamefont
  {Katz}, \citenamefont {Torsi},\ and\ \citenamefont
  {Haddon}}]{dodabalapur1995b}%
  \BibitemOpen
  \bibfield  {author} {\bibinfo {author} {\bibfnamefont {A.}~\bibnamefont
  {Dodabalapur}}, \bibinfo {author} {\bibfnamefont {H.~E.}\ \bibnamefont
  {Katz}}, \bibinfo {author} {\bibfnamefont {L.}~\bibnamefont {Torsi}}, \ and\
  \bibinfo {author} {\bibfnamefont {R.~C.}\ \bibnamefont {Haddon}},\ }\href
  {\doibase DOI: 10.1126/science.269.5230.1560} {\bibfield  {journal} {\bibinfo
   {journal} {Science}\ }\textbf {\bibinfo {volume} {269}},\ \bibinfo {pages}
  {1560} (\bibinfo {year} {1995}{\natexlab{b}})}\BibitemShut {NoStop}%
\bibitem [{\citenamefont {Hide}\ \emph {et~al.}(1996)\citenamefont {Hide},
  \citenamefont {Diaz-Garcia}, \citenamefont {Schwartz}, \citenamefont
  {Andersson}, \citenamefont {Pei},\ and\ \citenamefont {Heeger}}]{hide1996}%
  \BibitemOpen
  \bibfield  {author} {\bibinfo {author} {\bibfnamefont {F.}~\bibnamefont
  {Hide}}, \bibinfo {author} {\bibfnamefont {M.~A.}\ \bibnamefont
  {Diaz-Garcia}}, \bibinfo {author} {\bibfnamefont {B.~J.}\ \bibnamefont
  {Schwartz}}, \bibinfo {author} {\bibfnamefont {M.~R.}\ \bibnamefont
  {Andersson}}, \bibinfo {author} {\bibfnamefont {Q.~B.}\ \bibnamefont {Pei}},
  \ and\ \bibinfo {author} {\bibfnamefont {A.~J.}\ \bibnamefont {Heeger}},\
  }\href {\doibase DOI: 10.1126/science.273.5283.1833} {\bibfield  {journal}
  {\bibinfo  {journal} {Science}\ }\textbf {\bibinfo {volume} {273}},\ \bibinfo
  {pages} {1833} (\bibinfo {year} {1996})}\BibitemShut {NoStop}%
\bibitem [{\citenamefont {Yang}\ and\ \citenamefont {Swager}(1998)}]{yang1998}%
  \BibitemOpen
  \bibfield  {author} {\bibinfo {author} {\bibfnamefont {J.~S.}\ \bibnamefont
  {Yang}}\ and\ \bibinfo {author} {\bibfnamefont {T.~M.}\ \bibnamefont
  {Swager}},\ }\href@noop {} {\bibfield  {journal} {\bibinfo  {journal} {J. Am.
  Chem. Soc.}\ }\textbf {\bibinfo {volume} {120}},\ \bibinfo {pages} {5321}
  (\bibinfo {year} {1998})}\BibitemShut {NoStop}%
\bibitem [{\citenamefont {Schmitz}\ \emph {et~al.}(2001)\citenamefont
  {Schmitz}, \citenamefont {Posch}, \citenamefont {Thelakkat}, \citenamefont
  {Chmidt}, \citenamefont {Montali}, \citenamefont {Feldman}, \citenamefont
  {Smith},\ and\ \citenamefont {Weder}}]{schmitz2001}%
  \BibitemOpen
  \bibfield  {author} {\bibinfo {author} {\bibfnamefont {C.}~\bibnamefont
  {Schmitz}}, \bibinfo {author} {\bibfnamefont {P.}~\bibnamefont {Posch}},
  \bibinfo {author} {\bibfnamefont {M.}~\bibnamefont {Thelakkat}}, \bibinfo
  {author} {\bibfnamefont {H.~W.}\ \bibnamefont {Chmidt}}, \bibinfo {author}
  {\bibfnamefont {A.}~\bibnamefont {Montali}}, \bibinfo {author} {\bibfnamefont
  {K.}~\bibnamefont {Feldman}}, \bibinfo {author} {\bibfnamefont
  {P.}~\bibnamefont {Smith}}, \ and\ \bibinfo {author} {\bibfnamefont
  {C.}~\bibnamefont {Weder}},\ }\href {\doibase
  https://doi.org/10.1002/1616-3028(200102)11:1<41::AID-ADFM41>3.0.CO;2-S}
  {\bibfield  {journal} {\bibinfo  {journal} {Adv. Func. Mat.}\ }\textbf
  {\bibinfo {volume} {11}},\ \bibinfo {pages} {41} (\bibinfo {year}
  {2001})}\BibitemShut {NoStop}%
\bibitem [{\citenamefont {Nitzan}\ and\ \citenamefont
  {Ratner}(2003)}]{nitzan2003}%
  \BibitemOpen
  \bibfield  {author} {\bibinfo {author} {\bibfnamefont {A.}~\bibnamefont
  {Nitzan}}\ and\ \bibinfo {author} {\bibfnamefont {M.~A.}\ \bibnamefont
  {Ratner}},\ }\href {\doibase DOI: 10.1126/science.1081572} {\bibfield
  {journal} {\bibinfo  {journal} {Science}\ }\textbf {\bibinfo {volume}
  {300}},\ \bibinfo {pages} {1384} (\bibinfo {year} {2003})}\BibitemShut
  {NoStop}%
\bibitem [{\citenamefont {Jeckelmann}\ \emph {et~al.}(2000)\citenamefont
  {Jeckelmann}, \citenamefont {Gebhard},\ and\ \citenamefont
  {Essler}}]{jeckelmann2000}%
  \BibitemOpen
  \bibfield  {author} {\bibinfo {author} {\bibfnamefont {E.}~\bibnamefont
  {Jeckelmann}}, \bibinfo {author} {\bibfnamefont {F.}~\bibnamefont {Gebhard}},
  \ and\ \bibinfo {author} {\bibfnamefont {F.~H.~L.}\ \bibnamefont {Essler}},\
  }\href {\doibase https://doi.org/10.1103/PhysRevLett.85.3910} {\bibfield
  {journal} {\bibinfo  {journal} {Phys. Rev. lett.}\ }\textbf {\bibinfo
  {volume} {85}},\ \bibinfo {pages} {3910} (\bibinfo {year}
  {2000})}\BibitemShut {NoStop}%
\bibitem [{\citenamefont {Tsutsui}\ \emph {et~al.}(2000)\citenamefont
  {Tsutsui}, \citenamefont {Tohyama},\ and\ \citenamefont
  {Maekawa}}]{tsutsui2000}%
  \BibitemOpen
  \bibfield  {author} {\bibinfo {author} {\bibfnamefont {K.}~\bibnamefont
  {Tsutsui}}, \bibinfo {author} {\bibfnamefont {T.}~\bibnamefont {Tohyama}}, \
  and\ \bibinfo {author} {\bibfnamefont {S.}~\bibnamefont {Maekawa}},\ }\href
  {\doibase https://doi.org/10.1103/PhysRevB.61.7180} {\bibfield  {journal}
  {\bibinfo  {journal} {Phys. Rev. B}\ }\textbf {\bibinfo {volume} {61}},\
  \bibinfo {pages} {7180} (\bibinfo {year} {2000})}\BibitemShut {NoStop}%
\bibitem [{\citenamefont {Essler}\ \emph {et~al.}(2001)\citenamefont {Essler},
  \citenamefont {Gebhard},\ and\ \citenamefont {Jeckelmann}}]{essler2001}%
  \BibitemOpen
  \bibfield  {author} {\bibinfo {author} {\bibfnamefont {F.~H.~L.}\
  \bibnamefont {Essler}}, \bibinfo {author} {\bibfnamefont {F.}~\bibnamefont
  {Gebhard}}, \ and\ \bibinfo {author} {\bibfnamefont {E.}~\bibnamefont
  {Jeckelmann}},\ }\href {\doibase 10.1103/PhysRevB.64.125119} {\bibfield
  {journal} {\bibinfo  {journal} {Phys. Rev. B}\ }\textbf {\bibinfo {volume}
  {64}},\ \bibinfo {pages} {125119} (\bibinfo {year} {2001})}\BibitemShut
  {NoStop}%
\bibitem [{\citenamefont {Jeckelmann}(2003)}]{jeckelmann2003}%
  \BibitemOpen
  \bibfield  {author} {\bibinfo {author} {\bibfnamefont {E.}~\bibnamefont
  {Jeckelmann}},\ }\href {\doibase https://doi.org/10.1103/PhysRevB.67.075106}
  {\bibfield  {journal} {\bibinfo  {journal} {Phys. Rev. B}\ }\textbf {\bibinfo
  {volume} {67}},\ \bibinfo {pages} {075106} (\bibinfo {year}
  {2003})}\BibitemShut {NoStop}%
\bibitem [{\citenamefont {Gallagher}\ and\ \citenamefont
  {Mazumdar}(1997)}]{gallagher1997}%
  \BibitemOpen
  \bibfield  {author} {\bibinfo {author} {\bibfnamefont {F.~B.}\ \bibnamefont
  {Gallagher}}\ and\ \bibinfo {author} {\bibfnamefont {S.}~\bibnamefont
  {Mazumdar}},\ }\href {\doibase https://doi.org/10.1103/PhysRevB.56.15025}
  {\bibfield  {journal} {\bibinfo  {journal} {Phys. Rev. B}\ }\textbf {\bibinfo
  {volume} {56}},\ \bibinfo {pages} {15025} (\bibinfo {year}
  {1997})}\BibitemShut {NoStop}%
\bibitem [{\citenamefont {Barford}(2002)}]{barford2002}%
  \BibitemOpen
  \bibfield  {author} {\bibinfo {author} {\bibfnamefont {W.}~\bibnamefont
  {Barford}},\ }\href {\doibase 10.1103/PhysRevB.65.205118} {\bibfield
  {journal} {\bibinfo  {journal} {Phys. Rev. B}\ }\textbf {\bibinfo {volume}
  {65}},\ \bibinfo {pages} {205118} (\bibinfo {year} {2002})}\BibitemShut
  {NoStop}%
\bibitem [{\citenamefont {Gebhard}\ \emph {et~al.}(1997)\citenamefont
  {Gebhard}, \citenamefont {Born}, \citenamefont {Scheidler}, \citenamefont
  {Thomas},\ and\ \citenamefont {Koch}}]{Gebhard1997}%
  \BibitemOpen
  \bibfield  {author} {\bibinfo {author} {\bibfnamefont {F.}~\bibnamefont
  {Gebhard}}, \bibinfo {author} {\bibfnamefont {K.}~\bibnamefont {Born}},
  \bibinfo {author} {\bibfnamefont {M.}~\bibnamefont {Scheidler}}, \bibinfo
  {author} {\bibfnamefont {P.}~\bibnamefont {Thomas}}, \ and\ \bibinfo {author}
  {\bibfnamefont {S.~W.}\ \bibnamefont {Koch}},\ }\href {\doibase
  https://doi.org/10.1080/13642819708205701} {\bibfield  {journal} {\bibinfo
  {journal} {Phil. Mag. B}\ }\textbf {\bibinfo {volume} {75}},\ \bibinfo
  {pages} {46} (\bibinfo {year} {1997})}\BibitemShut {NoStop}%
\bibitem [{\citenamefont {Kancharla}\ and\ \citenamefont
  {Bolech}(2001)}]{Kancharla2001}%
  \BibitemOpen
  \bibfield  {author} {\bibinfo {author} {\bibfnamefont {S.~S.}\ \bibnamefont
  {Kancharla}}\ and\ \bibinfo {author} {\bibfnamefont {C.~J.}\ \bibnamefont
  {Bolech}},\ }\href {\doibase 10.1103/PhysRevB.64.085119} {\bibfield
  {journal} {\bibinfo  {journal} {Phys. Rev. B}\ }\textbf {\bibinfo {volume}
  {64}},\ \bibinfo {pages} {085119} (\bibinfo {year} {2001})}\BibitemShut
  {NoStop}%
\bibitem [{\citenamefont {Mizuno}\ \emph {et~al.}(2000)\citenamefont {Mizuno},
  \citenamefont {Tsutsui}, \citenamefont {Tohyama},\ and\ \citenamefont
  {Maekawa}}]{Mizuno2000}%
  \BibitemOpen
  \bibfield  {author} {\bibinfo {author} {\bibfnamefont {Y.}~\bibnamefont
  {Mizuno}}, \bibinfo {author} {\bibfnamefont {K.}~\bibnamefont {Tsutsui}},
  \bibinfo {author} {\bibfnamefont {T.}~\bibnamefont {Tohyama}}, \ and\
  \bibinfo {author} {\bibfnamefont {S.}~\bibnamefont {Maekawa}},\ }\href
  {\doibase 10.1103/PhysRevB.62.R4769} {\bibfield  {journal} {\bibinfo
  {journal} {Phys. Rev. B}\ }\textbf {\bibinfo {volume} {62}},\ \bibinfo
  {pages} {R4769} (\bibinfo {year} {2000})}\BibitemShut {NoStop}%
\bibitem [{\citenamefont {Glocke}\ \emph {et~al.}(2007)\citenamefont {Glocke},
  \citenamefont {Kl\"umper},\ and\ \citenamefont {Sirker}}]{Glocke2007}%
  \BibitemOpen
  \bibfield  {author} {\bibinfo {author} {\bibfnamefont {S.}~\bibnamefont
  {Glocke}}, \bibinfo {author} {\bibfnamefont {A.}~\bibnamefont {Kl\"umper}}, \
  and\ \bibinfo {author} {\bibfnamefont {J.}~\bibnamefont {Sirker}},\ }\href
  {\doibase 10.1103/PhysRevB.76.155121} {\bibfield  {journal} {\bibinfo
  {journal} {Phys. Rev. B}\ }\textbf {\bibinfo {volume} {76}},\ \bibinfo
  {pages} {155121} (\bibinfo {year} {2007})}\BibitemShut {NoStop}%
\bibitem [{\citenamefont {Matsueda}\ \emph {et~al.}(2004)\citenamefont
  {Matsueda}, \citenamefont {Tohyama},\ and\ \citenamefont
  {Maekawa}}]{Matsueda2008}%
  \BibitemOpen
  \bibfield  {author} {\bibinfo {author} {\bibfnamefont {H.}~\bibnamefont
  {Matsueda}}, \bibinfo {author} {\bibfnamefont {T.}~\bibnamefont {Tohyama}}, \
  and\ \bibinfo {author} {\bibfnamefont {S.}~\bibnamefont {Maekawa}},\ }\href
  {\doibase 10.1103/PhysRevB.70.033102} {\bibfield  {journal} {\bibinfo
  {journal} {Phys. Rev. B}\ }\textbf {\bibinfo {volume} {70}},\ \bibinfo
  {pages} {033102} (\bibinfo {year} {2004})}\BibitemShut {NoStop}%
\bibitem [{\citenamefont {Lu}\ \emph {et~al.}(2015)\citenamefont {Lu},
  \citenamefont {Shao}, \citenamefont {Bon\ifmmode~\check{c}\else \v{c}\fi{}a},
  \citenamefont {Manske},\ and\ \citenamefont {Tohyama}}]{Lu2015}%
  \BibitemOpen
  \bibfield  {author} {\bibinfo {author} {\bibfnamefont {H.}~\bibnamefont
  {Lu}}, \bibinfo {author} {\bibfnamefont {C.}~\bibnamefont {Shao}}, \bibinfo
  {author} {\bibfnamefont {J.}~\bibnamefont {Bon\ifmmode~\check{c}\else
  \v{c}\fi{}a}}, \bibinfo {author} {\bibfnamefont {D.}~\bibnamefont {Manske}},
  \ and\ \bibinfo {author} {\bibfnamefont {T.}~\bibnamefont {Tohyama}},\ }\href
  {\doibase 10.1103/PhysRevB.91.245117} {\bibfield  {journal} {\bibinfo
  {journal} {Phys. Rev. B}\ }\textbf {\bibinfo {volume} {91}},\ \bibinfo
  {pages} {245117} (\bibinfo {year} {2015})}\BibitemShut {NoStop}%
\bibitem [{\citenamefont {Al-Hassanieh}\ \emph {et~al.}(2008)\citenamefont
  {Al-Hassanieh}, \citenamefont {Reboredo}, \citenamefont {Feiguin},
  \citenamefont {Gonz\'alez},\ and\ \citenamefont {Dagotto}}]{exciton1}%
  \BibitemOpen
  \bibfield  {author} {\bibinfo {author} {\bibfnamefont {K.~A.}\ \bibnamefont
  {Al-Hassanieh}}, \bibinfo {author} {\bibfnamefont {F.~A.}\ \bibnamefont
  {Reboredo}}, \bibinfo {author} {\bibfnamefont {A.~E.}\ \bibnamefont
  {Feiguin}}, \bibinfo {author} {\bibfnamefont {I.}~\bibnamefont {Gonz\'alez}},
  \ and\ \bibinfo {author} {\bibfnamefont {E.}~\bibnamefont {Dagotto}},\ }\href
  {\doibase DOI: 10.1103/PhysRevLett.100.166403} {\bibfield  {journal}
  {\bibinfo  {journal} {Phys. Rev. Lett.}\ }\textbf {\bibinfo {volume} {100}},\
  \bibinfo {pages} {166403} (\bibinfo {year} {2008})}\BibitemShut {NoStop}%
\bibitem [{\citenamefont {Dias~da Silva}\ \emph {et~al.}(2010)\citenamefont
  {Dias~da Silva}, \citenamefont {Al-Hassanieh}, \citenamefont {Feiguin},
  \citenamefont {Reboredo},\ and\ \citenamefont {Dagotto}}]{exciton2}%
  \BibitemOpen
  \bibfield  {author} {\bibinfo {author} {\bibfnamefont {L.~G. G.~V.}\
  \bibnamefont {Dias~da Silva}}, \bibinfo {author} {\bibfnamefont {K.~A.}\
  \bibnamefont {Al-Hassanieh}}, \bibinfo {author} {\bibfnamefont {A.~E.}\
  \bibnamefont {Feiguin}}, \bibinfo {author} {\bibfnamefont {F.~A.}\
  \bibnamefont {Reboredo}}, \ and\ \bibinfo {author} {\bibfnamefont
  {E.}~\bibnamefont {Dagotto}},\ }\href {\doibase 10.1103/PhysRevB.81.125113}
  {\bibfield  {journal} {\bibinfo  {journal} {Phys. Rev. B}\ }\textbf {\bibinfo
  {volume} {81}},\ \bibinfo {pages} {125113} (\bibinfo {year}
  {2010})}\BibitemShut {NoStop}%
\bibitem [{\citenamefont {Rinc\'on}\ \emph {et~al.}(2014)\citenamefont
  {Rinc\'on}, \citenamefont {Al-Hassanieh}, \citenamefont {Feiguin},\ and\
  \citenamefont {Dagotto}}]{Rincon2014}%
  \BibitemOpen
  \bibfield  {author} {\bibinfo {author} {\bibfnamefont {J.}~\bibnamefont
  {Rinc\'on}}, \bibinfo {author} {\bibfnamefont {K.~A.}\ \bibnamefont
  {Al-Hassanieh}}, \bibinfo {author} {\bibfnamefont {A.~E.}\ \bibnamefont
  {Feiguin}}, \ and\ \bibinfo {author} {\bibfnamefont {E.}~\bibnamefont
  {Dagotto}},\ }\href {\doibase 10.1103/PhysRevB.90.155112} {\bibfield
  {journal} {\bibinfo  {journal} {Phys. Rev. B}\ }\textbf {\bibinfo {volume}
  {90}},\ \bibinfo {pages} {155112} (\bibinfo {year} {2014})}\BibitemShut
  {NoStop}%
\bibitem [{\citenamefont {Ono}\ \emph {et~al.}(2005)\citenamefont {Ono},
  \citenamefont {Kishida},\ and\ \citenamefont {Okamoto}}]{ono2005}%
  \BibitemOpen
  \bibfield  {author} {\bibinfo {author} {\bibfnamefont {M.}~\bibnamefont
  {Ono}}, \bibinfo {author} {\bibfnamefont {H.}~\bibnamefont {Kishida}}, \ and\
  \bibinfo {author} {\bibfnamefont {H.}~\bibnamefont {Okamoto}},\ }\href
  {\doibase https://doi.org/10.1103/PhysRevLett.95.087401} {\bibfield
  {journal} {\bibinfo  {journal} {Phys. Rev. Lett.}\ }\textbf {\bibinfo
  {volume} {95}},\ \bibinfo {pages} {087401} (\bibinfo {year}
  {2005})}\BibitemShut {NoStop}%
\bibitem [{\citenamefont {Schlappa}\ \emph {et~al.}(2012)\citenamefont
  {Schlappa}, \citenamefont {Wohlfeld}, \citenamefont {Zhou}, \citenamefont
  {Mourigal}, \citenamefont {Haverkort}, \citenamefont {Strocov}, \citenamefont
  {Hozoi}, \citenamefont {Monney}, \citenamefont {Nishimoto}, \citenamefont
  {Singh}, \citenamefont {Revcolevschi}, \citenamefont {Caux}, \citenamefont
  {Patthey}, \citenamefont {Ronnow}, \citenamefont {van~den Brink},\ and\
  \citenamefont {Schmitt}}]{Schlappa2012}%
  \BibitemOpen
  \bibfield  {author} {\bibinfo {author} {\bibfnamefont {J.}~\bibnamefont
  {Schlappa}}, \bibinfo {author} {\bibfnamefont {K.}~\bibnamefont {Wohlfeld}},
  \bibinfo {author} {\bibfnamefont {K.~J.}\ \bibnamefont {Zhou}}, \bibinfo
  {author} {\bibfnamefont {M.}~\bibnamefont {Mourigal}}, \bibinfo {author}
  {\bibfnamefont {M.~W.}\ \bibnamefont {Haverkort}}, \bibinfo {author}
  {\bibfnamefont {V.~N.}\ \bibnamefont {Strocov}}, \bibinfo {author}
  {\bibfnamefont {L.}~\bibnamefont {Hozoi}}, \bibinfo {author} {\bibfnamefont
  {C.}~\bibnamefont {Monney}}, \bibinfo {author} {\bibfnamefont
  {S.}~\bibnamefont {Nishimoto}}, \bibinfo {author} {\bibfnamefont
  {S.}~\bibnamefont {Singh}}, \bibinfo {author} {\bibfnamefont
  {A.}~\bibnamefont {Revcolevschi}}, \bibinfo {author} {\bibfnamefont {J.-S.}\
  \bibnamefont {Caux}}, \bibinfo {author} {\bibfnamefont {L.}~\bibnamefont
  {Patthey}}, \bibinfo {author} {\bibfnamefont {H.~M.}\ \bibnamefont {Ronnow}},
  \bibinfo {author} {\bibfnamefont {J.}~\bibnamefont {van~den Brink}}, \ and\
  \bibinfo {author} {\bibfnamefont {T.}~\bibnamefont {Schmitt}},\ }\href
  {\doibase https://doi.org/10.1038/nature10974} {\bibfield  {journal}
  {\bibinfo  {journal} {Nature}\ }\textbf {\bibinfo {volume} {485}},\ \bibinfo
  {pages} {82} (\bibinfo {year} {2012})}\BibitemShut {NoStop}%
\bibitem [{\citenamefont {Haldane}(1981)}]{Haldane1981}%
  \BibitemOpen
  \bibfield  {author} {\bibinfo {author} {\bibfnamefont {F.}~\bibnamefont
  {Haldane}},\ }\href {\doibase 10.1088/0022-3719/14/19/010} {\bibfield
  {journal} {\bibinfo  {journal} {J. of. Phys. C}\ }\textbf {\bibinfo {volume}
  {14}},\ \bibinfo {pages} {2585} (\bibinfo {year} {1981})}\BibitemShut
  {NoStop}%
\bibitem [{\citenamefont {Gogolin}\ \emph {et~al.}(1998)\citenamefont
  {Gogolin}, \citenamefont {Nersesyan},\ and\ \citenamefont
  {Tsvelik}}]{GogolinBook}%
  \BibitemOpen
  \bibfield  {author} {\bibinfo {author} {\bibfnamefont {A.~O.}\ \bibnamefont
  {Gogolin}}, \bibinfo {author} {\bibfnamefont {A.~A.}\ \bibnamefont
  {Nersesyan}}, \ and\ \bibinfo {author} {\bibfnamefont {A.~M.}\ \bibnamefont
  {Tsvelik}},\ }\href@noop {} {\emph {\bibinfo {title} {Bosonization of
  Strongly Correlated Systems}}}\ (\bibinfo  {publisher} {Cambridge University
  Press, Cambridge, England},\ \bibinfo {year} {1998})\BibitemShut {NoStop}%
\bibitem [{\citenamefont {Giamarchi}(2004)}]{GiamarchiBook}%
  \BibitemOpen
  \bibfield  {author} {\bibinfo {author} {\bibfnamefont {T.}~\bibnamefont
  {Giamarchi}},\ }\href@noop {} {\emph {\bibinfo {title} {Quantum Physics in
  One Dimension}}}\ (\bibinfo  {publisher} {Clarendon Press, Oxford},\ \bibinfo
  {year} {2004})\BibitemShut {NoStop}%
\bibitem [{\citenamefont {White}\ and\ \citenamefont
  {Feiguin}(2004)}]{White2004a}%
  \BibitemOpen
  \bibfield  {author} {\bibinfo {author} {\bibfnamefont {S.~R.}\ \bibnamefont
  {White}}\ and\ \bibinfo {author} {\bibfnamefont {A.~E.}\ \bibnamefont
  {Feiguin}},\ }\href {\doibase 10.1103/PhysRevLett.93.076401} {\bibfield
  {journal} {\bibinfo  {journal} {Phys. Rev. Lett.}\ }\textbf {\bibinfo
  {volume} {93}},\ \bibinfo {pages} {076401} (\bibinfo {year}
  {2004})}\BibitemShut {NoStop}%
\bibitem [{\citenamefont {Daley}\ \emph {et~al.}(2004)\citenamefont {Daley},
  \citenamefont {Kollath}, \citenamefont {Schollw\"ock},\ and\ \citenamefont
  {Vidal}}]{Daley2004}%
  \BibitemOpen
  \bibfield  {author} {\bibinfo {author} {\bibfnamefont {A.~J.}\ \bibnamefont
  {Daley}}, \bibinfo {author} {\bibfnamefont {C.}~\bibnamefont {Kollath}},
  \bibinfo {author} {\bibfnamefont {U.}~\bibnamefont {Schollw\"ock}}, \ and\
  \bibinfo {author} {\bibfnamefont {G.}~\bibnamefont {Vidal}},\ }\href
  {\doibase 10.1088/1742-5468/2004/04/p04005} {\bibfield  {journal} {\bibinfo
  {journal} {Journal of Statistical Mechanics: Theory and Experiment}\ }\textbf
  {\bibinfo {volume} {2004}},\ \bibinfo {pages} {P04005} (\bibinfo {year}
  {2004})}\BibitemShut {NoStop}%
\bibitem [{\citenamefont {Feiguin}(2011)}]{vietri}%
  \BibitemOpen
  \bibfield  {author} {\bibinfo {author} {\bibfnamefont {A.~E.}\ \bibnamefont
  {Feiguin}},\ }in\ \href@noop {} {\emph {\bibinfo {booktitle} {XV Training
  Course in the Physics of Strongly Correlated Systems}}},\ Vol.\ \bibinfo
  {volume} {1419}\ (\bibinfo  {publisher} {AIP Proceedings},\ \bibinfo {year}
  {2011})\ p.~\bibinfo {pages} {5}\BibitemShut {NoStop}%
\bibitem [{\citenamefont {Paeckel}\ \emph {et~al.}(2019)\citenamefont
  {Paeckel}, \citenamefont {Köhler}, \citenamefont {Swoboda}, \citenamefont
  {Manmana}, \citenamefont {Schollw\"ock},\ and\ \citenamefont
  {Hubig}}]{Paeckel2019}%
  \BibitemOpen
  \bibfield  {author} {\bibinfo {author} {\bibfnamefont {S.}~\bibnamefont
  {Paeckel}}, \bibinfo {author} {\bibfnamefont {T.}~\bibnamefont {Köhler}},
  \bibinfo {author} {\bibfnamefont {A.}~\bibnamefont {Swoboda}}, \bibinfo
  {author} {\bibfnamefont {S.~R.}\ \bibnamefont {Manmana}}, \bibinfo {author}
  {\bibfnamefont {U.}~\bibnamefont {Schollw\"ock}}, \ and\ \bibinfo {author}
  {\bibfnamefont {C.}~\bibnamefont {Hubig}},\ }\href@noop {} {\  (\bibinfo
  {year} {2019})},\ \Eprint {http://arxiv.org/abs/1901.05824}
  {arXiv:1901.05824} \BibitemShut {NoStop}%
\bibitem [{\citenamefont {Nakamura}(2000)}]{Nakamura2000}%
  \BibitemOpen
  \bibfield  {author} {\bibinfo {author} {\bibfnamefont {M.}~\bibnamefont
  {Nakamura}},\ }\href {\doibase 10.1103/PhysRevB.61.16377} {\bibfield
  {journal} {\bibinfo  {journal} {Phys. Rev. B}\ }\textbf {\bibinfo {volume}
  {61}},\ \bibinfo {pages} {16377} (\bibinfo {year} {2000})}\BibitemShut
  {NoStop}%
\bibitem [{\citenamefont {Jeckelmann}(2002)}]{Jeckelmann2002}%
  \BibitemOpen
  \bibfield  {author} {\bibinfo {author} {\bibfnamefont {E.}~\bibnamefont
  {Jeckelmann}},\ }\href {\doibase https://doi.org/10.1103/PhysRevB.66.045114}
  {\bibfield  {journal} {\bibinfo  {journal} {Phys. Rev. B}\ }\textbf {\bibinfo
  {volume} {66}},\ \bibinfo {pages} {045114} (\bibinfo {year}
  {2002})}\BibitemShut {NoStop}%
\bibitem [{\citenamefont {Sandvik}\ \emph {et~al.}(2003)\citenamefont
  {Sandvik}, \citenamefont {Sengupta},\ and\ \citenamefont
  {Campbell}}]{Sandvik2003}%
  \BibitemOpen
  \bibfield  {author} {\bibinfo {author} {\bibfnamefont {A.~W.}\ \bibnamefont
  {Sandvik}}, \bibinfo {author} {\bibfnamefont {P.}~\bibnamefont {Sengupta}}, \
  and\ \bibinfo {author} {\bibfnamefont {D.~K.}\ \bibnamefont {Campbell}},\
  }\href {\doibase 10.1103/PhysRevLett.91.089701} {\bibfield  {journal}
  {\bibinfo  {journal} {Phys. Rev. Lett.}\ }\textbf {\bibinfo {volume} {91}},\
  \bibinfo {pages} {089701} (\bibinfo {year} {2003})}\BibitemShut {NoStop}%
\bibitem [{\citenamefont {Tsuchiizu}\ and\ \citenamefont
  {Furusaki}(2004)}]{Tsuchiizu2004}%
  \BibitemOpen
  \bibfield  {author} {\bibinfo {author} {\bibfnamefont {M.}~\bibnamefont
  {Tsuchiizu}}\ and\ \bibinfo {author} {\bibfnamefont {A.}~\bibnamefont
  {Furusaki}},\ }\href {\doibase 10.1103/PhysRevB.69.035103} {\bibfield
  {journal} {\bibinfo  {journal} {Phys. Rev. B}\ }\textbf {\bibinfo {volume}
  {69}},\ \bibinfo {pages} {035103} (\bibinfo {year} {2004})}\BibitemShut
  {NoStop}%
\bibitem [{\citenamefont {Sensarma}\ \emph {et~al.}(2010)\citenamefont
  {Sensarma}, \citenamefont {Pekker}, \citenamefont {Altman}, \citenamefont
  {Demler}, \citenamefont {Strohmaier}, \citenamefont {Greif}, \citenamefont
  {J\"ordens}, \citenamefont {Tarruell}, \citenamefont {Moritz},\ and\
  \citenamefont {Esslinger}}]{Sensarma2010}%
  \BibitemOpen
  \bibfield  {author} {\bibinfo {author} {\bibfnamefont {R.}~\bibnamefont
  {Sensarma}}, \bibinfo {author} {\bibfnamefont {D.}~\bibnamefont {Pekker}},
  \bibinfo {author} {\bibfnamefont {E.}~\bibnamefont {Altman}}, \bibinfo
  {author} {\bibfnamefont {E.}~\bibnamefont {Demler}}, \bibinfo {author}
  {\bibfnamefont {N.}~\bibnamefont {Strohmaier}}, \bibinfo {author}
  {\bibfnamefont {D.}~\bibnamefont {Greif}}, \bibinfo {author} {\bibfnamefont
  {R.}~\bibnamefont {J\"ordens}}, \bibinfo {author} {\bibfnamefont
  {L.}~\bibnamefont {Tarruell}}, \bibinfo {author} {\bibfnamefont
  {H.}~\bibnamefont {Moritz}}, \ and\ \bibinfo {author} {\bibfnamefont
  {T.}~\bibnamefont {Esslinger}},\ }\href {\doibase 10.1103/PhysRevB.82.224302}
  {\bibfield  {journal} {\bibinfo  {journal} {Phys. Rev. B}\ }\textbf {\bibinfo
  {volume} {82}},\ \bibinfo {pages} {224302} (\bibinfo {year}
  {2010})}\BibitemShut {NoStop}%
\bibitem [{\citenamefont {Eckstein}\ and\ \citenamefont
  {Werner}(2011)}]{Eckstein2011}%
  \BibitemOpen
  \bibfield  {author} {\bibinfo {author} {\bibfnamefont {M.}~\bibnamefont
  {Eckstein}}\ and\ \bibinfo {author} {\bibfnamefont {P.}~\bibnamefont
  {Werner}},\ }\href {\doibase 10.1103/PhysRevB.84.035122} {\bibfield
  {journal} {\bibinfo  {journal} {Phys. Rev. B}\ }\textbf {\bibinfo {volume}
  {84}},\ \bibinfo {pages} {035122} (\bibinfo {year} {2011})}\BibitemShut
  {NoStop}%
\bibitem [{\citenamefont {Lenar\ifmmode \check{c}\else
  \v{c}\fi{}i\ifmmode~\check{c}\else \v{c}\fi{}}\ and\ \citenamefont
  {Prelov\ifmmode~\check{s}\else \v{s}\fi{}ek}(2013)}]{Lenarcic2013}%
  \BibitemOpen
  \bibfield  {author} {\bibinfo {author} {\bibfnamefont {Z.}~\bibnamefont
  {Lenar\ifmmode \check{c}\else \v{c}\fi{}i\ifmmode~\check{c}\else
  \v{c}\fi{}}}\ and\ \bibinfo {author} {\bibfnamefont {P.}~\bibnamefont
  {Prelov\ifmmode~\check{s}\else \v{s}\fi{}ek}},\ }\href {\doibase
  10.1103/PhysRevLett.111.016401} {\bibfield  {journal} {\bibinfo  {journal}
  {Phys. Rev. Lett.}\ }\textbf {\bibinfo {volume} {111}},\ \bibinfo {pages}
  {016401} (\bibinfo {year} {2013})}\BibitemShut {NoStop}%
\bibitem [{\citenamefont {Eckstein}\ and\ \citenamefont
  {Werner}(2016)}]{Eckstein2016}%
  \BibitemOpen
  \bibfield  {author} {\bibinfo {author} {\bibfnamefont {M.}~\bibnamefont
  {Eckstein}}\ and\ \bibinfo {author} {\bibfnamefont {P.}~\bibnamefont
  {Werner}},\ }\href {https://doi.org/10.1038/srep21235} {\bibfield  {journal}
  {\bibinfo  {journal} {Scientific Reports}\ }\textbf {\bibinfo {volume} {6}},\
  \bibinfo {pages} {21235 EP } (\bibinfo {year} {2016})},\ \bibinfo {note}
  {article L3 -}\BibitemShut {NoStop}%
\bibitem [{\citenamefont {Feiguin}\ and\ \citenamefont
  {White}(2005)}]{Feiguin2005a}%
  \BibitemOpen
  \bibfield  {author} {\bibinfo {author} {\bibfnamefont {A.}~\bibnamefont
  {Feiguin}}\ and\ \bibinfo {author} {\bibfnamefont {S.}~\bibnamefont
  {White}},\ }\href {\doibase https://doi.org/10.1103/PhysRevB.72.220401}
  {\bibfield  {journal} {\bibinfo  {journal} {Phys. Rev. B}\ }\textbf {\bibinfo
  {volume} {72}},\ \bibinfo {pages} {220401} (\bibinfo {year}
  {2005})}\BibitemShut {NoStop}%
\bibitem [{\citenamefont {Cheianov}\ and\ \citenamefont
  {Zvonarev}(2004)}]{Cheianov2004}%
  \BibitemOpen
  \bibfield  {author} {\bibinfo {author} {\bibfnamefont {V.~V.}\ \bibnamefont
  {Cheianov}}\ and\ \bibinfo {author} {\bibfnamefont {M.~B.}\ \bibnamefont
  {Zvonarev}},\ }\href {\doibase 10.1103/PhysRevLett.92.176401} {\bibfield
  {journal} {\bibinfo  {journal} {Phys. Rev. Lett.}\ }\textbf {\bibinfo
  {volume} {92}},\ \bibinfo {pages} {176401} (\bibinfo {year}
  {2004})}\BibitemShut {NoStop}%
\bibitem [{\citenamefont {Cheianov}\ \emph {et~al.}(2005)\citenamefont
  {Cheianov}, \citenamefont {Smith},\ and\ \citenamefont
  {Zvonarev}}]{Cheianov2005}%
  \BibitemOpen
  \bibfield  {author} {\bibinfo {author} {\bibfnamefont {V.~V.}\ \bibnamefont
  {Cheianov}}, \bibinfo {author} {\bibfnamefont {H.}~\bibnamefont {Smith}}, \
  and\ \bibinfo {author} {\bibfnamefont {M.~B.}\ \bibnamefont {Zvonarev}},\
  }\href {\doibase 10.1103/PhysRevA.71.033610} {\bibfield  {journal} {\bibinfo
  {journal} {Phys. Rev. A}\ }\textbf {\bibinfo {volume} {71}},\ \bibinfo
  {pages} {033610} (\bibinfo {year} {2005})}\BibitemShut {NoStop}%
\bibitem [{\citenamefont {Abendschein}\ and\ \citenamefont
  {Assaad}(2006)}]{Abendschein2006}%
  \BibitemOpen
  \bibfield  {author} {\bibinfo {author} {\bibfnamefont {A.}~\bibnamefont
  {Abendschein}}\ and\ \bibinfo {author} {\bibfnamefont {F.~F.}\ \bibnamefont
  {Assaad}},\ }\href {\doibase 10.1103/PhysRevB.73.165119} {\bibfield
  {journal} {\bibinfo  {journal} {Phys. Rev. B}\ }\textbf {\bibinfo {volume}
  {73}},\ \bibinfo {pages} {165119} (\bibinfo {year} {2006})}\BibitemShut
  {NoStop}%
\bibitem [{\citenamefont {Fiete}(2007)}]{Fiete2007b}%
  \BibitemOpen
  \bibfield  {author} {\bibinfo {author} {\bibfnamefont {G.}~\bibnamefont
  {Fiete}},\ }\href {\doibase https://doi.org/10.1103/RevModPhys.79.801}
  {\bibfield  {journal} {\bibinfo  {journal} {Rev. Mod. Phys.}\ }\textbf
  {\bibinfo {volume} {79}},\ \bibinfo {pages} {801} (\bibinfo {year}
  {2007})}\BibitemShut {NoStop}%
\bibitem [{\citenamefont {Halperin}(2007)}]{Halperin2007}%
  \BibitemOpen
  \bibfield  {author} {\bibinfo {author} {\bibfnamefont {B.~I.}\ \bibnamefont
  {Halperin}},\ }\href {\doibase https://doi.org/10.1063/1.2722722} {\bibfield
  {journal} {\bibinfo  {journal} {J. Appl. Phys.}\ }\textbf {\bibinfo {volume}
  {101}},\ \bibinfo {pages} {081601} (\bibinfo {year} {2007})}\BibitemShut
  {NoStop}%
\bibitem [{\citenamefont {Feiguin}\ and\ \citenamefont
  {Fiete}(2010)}]{Feiguin2009}%
  \BibitemOpen
  \bibfield  {author} {\bibinfo {author} {\bibfnamefont {A.~E.}\ \bibnamefont
  {Feiguin}}\ and\ \bibinfo {author} {\bibfnamefont {G.~A.}\ \bibnamefont
  {Fiete}},\ }\href {\doibase 10.1103/PhysRevB.81.075108} {\bibfield  {journal}
  {\bibinfo  {journal} {Phys. Rev. B}\ }\textbf {\bibinfo {volume} {81}},\
  \bibinfo {pages} {075108} (\bibinfo {year} {2010})}\BibitemShut {NoStop}%
\bibitem [{\citenamefont {Feiguin}\ and\ \citenamefont
  {Fiete}(2011)}]{Feiguin2011}%
  \BibitemOpen
  \bibfield  {author} {\bibinfo {author} {\bibfnamefont {A.~E.}\ \bibnamefont
  {Feiguin}}\ and\ \bibinfo {author} {\bibfnamefont {G.~A.}\ \bibnamefont
  {Fiete}},\ }\href {\doibase 10.1103/PhysRevLett.106.146401} {\bibfield
  {journal} {\bibinfo  {journal} {Phys. Rev. Lett.}\ }\textbf {\bibinfo
  {volume} {106}},\ \bibinfo {pages} {146401} (\bibinfo {year}
  {2011})}\BibitemShut {NoStop}%
\bibitem [{\citenamefont {Soltanieh-ha}\ and\ \citenamefont
  {Feiguin}(2014)}]{Soltanieh-ha2014}%
  \BibitemOpen
  \bibfield  {author} {\bibinfo {author} {\bibfnamefont {M.}~\bibnamefont
  {Soltanieh-ha}}\ and\ \bibinfo {author} {\bibfnamefont {A.~E.}\ \bibnamefont
  {Feiguin}},\ }\href {\doibase 10.1103/PhysRevB.90.165145} {\bibfield
  {journal} {\bibinfo  {journal} {Phys. Rev. B}\ }\textbf {\bibinfo {volume}
  {90}},\ \bibinfo {pages} {165145} (\bibinfo {year} {2014})}\BibitemShut
  {NoStop}%
\bibitem [{\citenamefont {Nocera}\ \emph {et~al.}(2018)\citenamefont {Nocera},
  \citenamefont {Essler},\ and\ \citenamefont {Feiguin}}]{Nocera2018}%
  \BibitemOpen
  \bibfield  {author} {\bibinfo {author} {\bibfnamefont {A.}~\bibnamefont
  {Nocera}}, \bibinfo {author} {\bibfnamefont {F.~H.~L.}\ \bibnamefont
  {Essler}}, \ and\ \bibinfo {author} {\bibfnamefont {A.~E.}\ \bibnamefont
  {Feiguin}},\ }\href {\doibase 10.1103/PhysRevB.97.045146} {\bibfield
  {journal} {\bibinfo  {journal} {Phys. Rev. B}\ }\textbf {\bibinfo {volume}
  {97}},\ \bibinfo {pages} {045146} (\bibinfo {year} {2018})}\BibitemShut
  {NoStop}%
\bibitem [{\citenamefont {Kidd}\ \emph {et~al.}(2008)\citenamefont {Kidd},
  \citenamefont {Valla}, \citenamefont {Johnson}, \citenamefont {Kim},
  \citenamefont {Gu},\ and\ \citenamefont {Homes}}]{kidd08}%
  \BibitemOpen
  \bibfield  {author} {\bibinfo {author} {\bibfnamefont {T.~E.}\ \bibnamefont
  {Kidd}}, \bibinfo {author} {\bibfnamefont {T.}~\bibnamefont {Valla}},
  \bibinfo {author} {\bibfnamefont {P.~D.}\ \bibnamefont {Johnson}}, \bibinfo
  {author} {\bibfnamefont {K.~W.}\ \bibnamefont {Kim}}, \bibinfo {author}
  {\bibfnamefont {G.~D.}\ \bibnamefont {Gu}}, \ and\ \bibinfo {author}
  {\bibfnamefont {C.~C.}\ \bibnamefont {Homes}},\ }\href {\doibase
  10.1103/PhysRevB.77.054503} {\bibfield  {journal} {\bibinfo  {journal} {Phys.
  Rev. B}\ }\textbf {\bibinfo {volume} {77}},\ \bibinfo {pages} {054503}
  (\bibinfo {year} {2008})}\BibitemShut {NoStop}%
\bibitem [{\citenamefont {Kobayashi}\ \emph {et~al.}(1999)\citenamefont
  {Kobayashi}, \citenamefont {Mizokawa}, \citenamefont {Fujimori},
  \citenamefont {Isobe}, \citenamefont {Ueda}, \citenamefont {Tohyama},\ and\
  \citenamefont {Maekawa}}]{maekawa99}%
  \BibitemOpen
  \bibfield  {author} {\bibinfo {author} {\bibfnamefont {K.}~\bibnamefont
  {Kobayashi}}, \bibinfo {author} {\bibfnamefont {T.}~\bibnamefont {Mizokawa}},
  \bibinfo {author} {\bibfnamefont {A.}~\bibnamefont {Fujimori}}, \bibinfo
  {author} {\bibfnamefont {M.}~\bibnamefont {Isobe}}, \bibinfo {author}
  {\bibfnamefont {Y.}~\bibnamefont {Ueda}}, \bibinfo {author} {\bibfnamefont
  {T.}~\bibnamefont {Tohyama}}, \ and\ \bibinfo {author} {\bibfnamefont
  {S.}~\bibnamefont {Maekawa}},\ }\href {\doibase 10.1103/PhysRevLett.82.803}
  {\bibfield  {journal} {\bibinfo  {journal} {Phys. Rev. Lett.}\ }\textbf
  {\bibinfo {volume} {82}},\ \bibinfo {pages} {803} (\bibinfo {year}
  {1999})}\BibitemShut {NoStop}%
\bibitem [{\citenamefont {Lu}\ \emph {et~al.}(2012)\citenamefont {Lu},
  \citenamefont {Sota}, \citenamefont {Matsueda}, \citenamefont {Bon\v{c}a},\
  and\ \citenamefont {Tohyama}}]{Lu2012a}%
  \BibitemOpen
  \bibfield  {author} {\bibinfo {author} {\bibfnamefont {H.}~\bibnamefont
  {Lu}}, \bibinfo {author} {\bibfnamefont {S.}~\bibnamefont {Sota}}, \bibinfo
  {author} {\bibfnamefont {H.}~\bibnamefont {Matsueda}}, \bibinfo {author}
  {\bibfnamefont {J.}~\bibnamefont {Bon\v{c}a}}, \ and\ \bibinfo {author}
  {\bibfnamefont {T.}~\bibnamefont {Tohyama}},\ }\href {\doibase
  10.1103/PhysRevLett.109.197401} {\bibfield  {journal} {\bibinfo  {journal}
  {Phys. Rev. Lett.}\ }\textbf {\bibinfo {volume} {109}},\ \bibinfo {pages}
  {197401} (\bibinfo {year} {2012})}\BibitemShut {NoStop}%
\end{thebibliography}%
\end{document}